\documentclass[aps,prl,amsmath,amssymb,twocolumn,superscriptaddress,longbibliography]{revtex4-1}
\usepackage{graphicx}
\usepackage{braket}
\usepackage{color}
\usepackage{natbib}
\usepackage{hyperref}
\usepackage{braket}
\usepackage{color}

\begin{document}


\title{Spectrum of the Nuclear Environment for GaAs Spin Qubits}

\author{Filip~K.~Malinowski}
\affiliation{Center for Quantum Devices, Niels Bohr Institute, University of Copenhagen, 2100 Copenhagen, Denmark}

\author{Frederico~Martins}
\affiliation{Center for Quantum Devices, Niels Bohr Institute, University of Copenhagen, 2100 Copenhagen, Denmark}

\author{{\L}ukasz Cywi\'nski}
\affiliation{Institute of Physics, Polish Academy of Sciences, Aleja Lotnikow 32/46, PL-02668 Warsaw, Poland}

\author{Mark~S.~Rudner}
\affiliation{Center for Quantum Devices, Niels Bohr Institute, University of Copenhagen, 2100 Copenhagen, Denmark}
\affiliation{Niels Bohr International Academy, Niels Bohr Institute, 2100 Copenhagen, Denmark}

\author{Peter~D.~Nissen}
\affiliation{Center for Quantum Devices, Niels Bohr Institute, University of Copenhagen, 2100 Copenhagen, Denmark}

\author{Saeed~Fallahi}
\affiliation{Department of Physics and Astronomy, Birck Nanotechnology Center, Purdue University, West Lafayette, Indiana 47907, USA}

\author{Geoffrey~C.~Gardner}
\affiliation{Department of Physics and Astronomy, Birck Nanotechnology Center, Purdue University, West Lafayette, Indiana 47907, USA}
\affiliation{School of Materials Engineering and School of Electrical and Computer Engineering, Purdue University, West Lafayette, Indiana 47907, USA}

\author{Michael~J.~Manfra}
\affiliation{Department of Physics and Astronomy, Birck Nanotechnology Center, and Station Q Purdue, Purdue University, West Lafayette, Indiana 47907, USA}
\affiliation{School of Materials Engineering, Purdue University, West Lafayette, Indiana 47907, USA}

\author{Charles~M.~Marcus}
\affiliation{Center for Quantum Devices and Station Q Copenhagen, Niels Bohr Institute, University of Copenhagen, 2100 Copenhagen, Denmark}

\author{Ferdinand~Kuemmeth}
\affiliation{Center for Quantum Devices, Niels Bohr Institute, University of Copenhagen, 2100 Copenhagen, Denmark}

\newcommand{\VL}{V_\mathrm{L}}
\newcommand{\VM}{V_\mathrm{M}}
\newcommand{\VR}{V_\mathrm{R}}

\newcommand{\Btot}{B^\mathrm{tot}}
\newcommand{\Bext}{B_\mathrm{ext}}
\newcommand{\Bznuc}{B_\mathrm{z}^\mathrm{nuc}}
\newcommand{\Bpnuc}{B_\perp^\mathrm{nuc}}

\newcommand{\ud}{\uparrow\downarrow}
\newcommand{\du}{\downarrow\uparrow}

\newcommand{\drv}{\mathrm{d}}

\newcommand{\Ga}{^{69}\mathrm{Ga}}
\newcommand{\Gb}{^{71}\mathrm{Ga}}
\newcommand{\As}{^{75}\mathrm{As}}
\newcommand{\fGa}{f_{^{69}{\rm Ga}}}
\newcommand{\fGb}{f_{^{71}{\rm Ga}}}
\newcommand{\fAs}{f_{^{75}{\rm As}}}

\newcommand{\TCPMG}{T_2 ^\mathrm{CPMG}}
\renewcommand{\vec}[1]{{\bf #1}}
\newcommand{\Tn}{T_{2,n}}

\newcommand{\MatEl}[3]{\langle \, #1 \,\vert\,#2\,\vert\,#3\,\rangle}
\newcommand{\Amp}[2]{\langle \, #1\, \vert\,  #2 \, \rangle}
\newcommand{\mpar}[1]{\marginpar{\small \it #1}}
\newcommand{\Avg}[1]{\langle  #1  \rangle}
\newcommand{\la}{\langle}
\newcommand{\ra}{\rangle}
\newcommand{\lb}{\left[}
\newcommand{\rb}{\right]}
\newcommand{\lp}{\left(}
\newcommand{\rp}{\right)}
\newcommand{\E}{{\cal E}}
\newcommand{\HH}{{\cal H}}
\newcommand{\LL}{{\cal L}}
\newcommand{\tr}{{\rm tr}\,}
\newcommand{\p}{\partial}

\date{\today}

\begin{abstract}
	Using a singlet-triplet spin qubit as a sensitive spectrometer of the GaAs nuclear spin bath, we demonstrate that the  spectrum of Overhauser noise agrees with a classical spin diffusion model over six orders of magnitude in frequency, from 1~mHz to 1~kHz, is flat below 10~mHz, and falls as $1/f^2$ for frequency $f \! \gtrsim \! 1$~Hz. Increasing the applied magnetic field from 0.1~T to 0.75~T suppresses electron-mediated spin diffusion, which decreases spectral content in the $1/f^2$ region and lowers the saturation frequency, each by an order of magnitude, consistent with a numerical model. Spectral content at megahertz frequencies is accessed using dynamical decoupling, which shows a crossover from the few-pulse regime ($\lesssim \! 16~\pi$-pulses), where transverse Overhauser fluctuations dominate dephasing, to the many-pulse regime ($\gtrsim \! 32$~$\pi$-pulses), where  longitudinal Overhauser fluctuations with a $1/f$ spectrum dominate. 
\end{abstract}

\maketitle 

Precise control of single electron spins in gate-defined quantum dots makes them a promising platform for quantum computation~\cite{Loss1998,Veldhorst2015,Petta2005,Nowack2011,Shulman2012}.
In particular, GaAs spin qubits benefit from unmatched reliability in fabrication and tuning.
However, being a III-V semiconductor, the GaAs lattice hosts spinful nuclei that couple to electron spins via the hyperfine interaction~\cite{Petta2005,Shulman2012,Malinowski2017,Bluhm2011,Foletti2010}.
Nuclear dynamics lead to fluctuations of the Overhauser field, which affect the coherent evolution of spin qubits. 
In turn, advances in qubit operation, including single-shot readout \cite{Barthel2009} and long dynamical decoupling sequences \cite{Malinowski2017}, allow spin qubits to serve as sensitive probes of the electron-plus-nuclear-environment system, an interesting coupled nonlinear many-body system.

\begin{figure}[hbt]
	\includegraphics[width=0.47\textwidth]{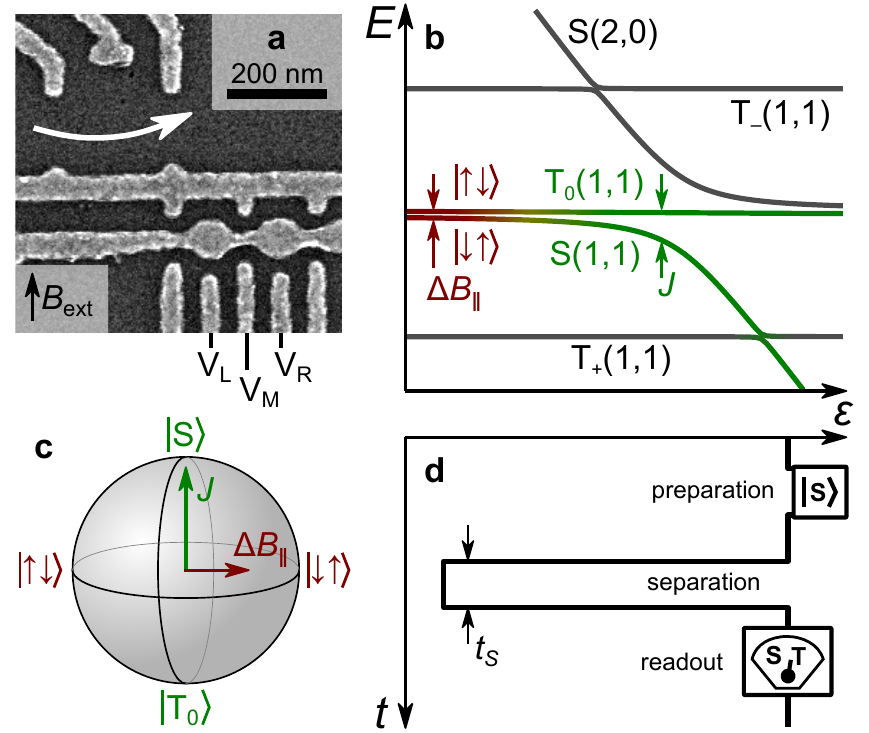}
	\caption{(a) Electron micrograph of the device. Gate voltages  $V_i$ control the double dot state on ns timescales. Reflectance from the RF resonant circuit incorporating a sensor dot (white arrow) measures the charge state of the double dot located below the round accumulation gates.
	(b) Energy levels of the two-electron double dot as a function of detuning $\varepsilon \! = \! V_L-V_R$ at the (1,1)-(2,0) charge transition. Red-green lines indicate the qubit states.
	(c) Bloch sphere representation of the qubit. Rotation axes correspond to exchange interaction $J$ (green) and gradient of the Overhauser field $\Delta B_\parallel$ (red).
	(d) Pulse cycle used to probe the qubit precession in the gradient of the Overhauser field. The qubit is initialized in the S(2,0) state by exchanging electrons with the lead. Next, one electron is moved to the right dot, and the qubit evolves for the time $t_S$ in the gradient of the Overhauser field. Finally, $\varepsilon$ is pulsed back to the readout point, projecting $\ket{S}$ into a (2,0) charge state, whereas $\ket{T_0}$ remains in (1,1). 
	}
	\label{fig1}
\end{figure}

In this Letter, we use a singlet-triplet (S-T$_0$) qubit as a probe to reveal the dynamics and magnetic field dependence of the GaAs nuclear spin bath over a wide range of frequencies, without the use of nuclear pumping~\cite{Shulman2014,Bechtold2015,Nichol2015}  or postselection~\cite{Delbecq2016} techniques. 
The qubit is defined in a two-electron double quantum dot (Fig.~\ref{fig1}a). The external magnetic field $\Bext$ separates the qubit states singlet, $\ket{\mathrm{S}} \! = \! \tfrac{1}{\sqrt{2}}(\ket{\ud} \! - \! \ket{\du})$, and the unpolarized triplet, $\ket{\mathrm{T_0}} \! = \! \tfrac{1}{\sqrt{2}}(\ket{\ud} \! + \! \ket{\du})$, from the fully polarized triplet states, $\ket{T_+} \! = \! \ket{\uparrow\uparrow}$ and $\ket{T_-} \! = \! \ket{\downarrow\downarrow}$. In this notation, the first (second) arrow indicates the spin in the left (right) dot. 
The resulting energy diagram of the spin states at the transition between (1,1) and (2,0) charge states is presented in Fig.~\ref{fig1}b.
Here ($N$,$M$) indicates the number of electrons in the left ($N$) and the right ($M$) dot. 
The Bloch sphere representation of the qubit is shown in Fig.~\ref{fig1}c.

\begin{figure}[hbt]
	\includegraphics[scale=0.9]{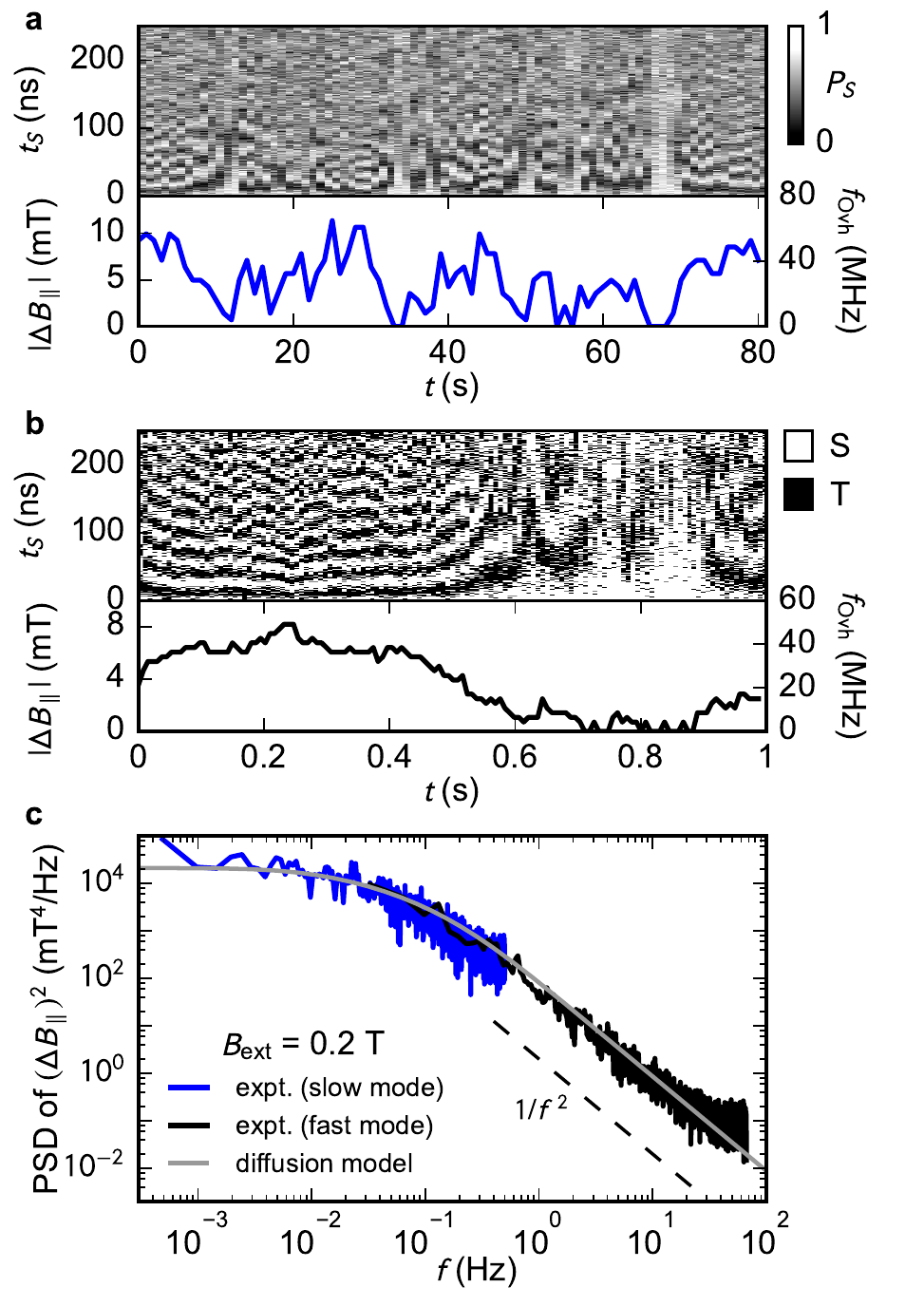}
	\caption{(a,b) Top panels present S-T$_0$ oscillations resulting from the relative precession of the two electron spins in the Overhauser field gradient, as a function of laboratory time at $\Bext \! = \! 0.2$~T (see main text). In the bottom panels we show the extracted frequency of oscillations, $f_\mathrm{Ovh}$, converted to $|\Delta B_\parallel|$.
	(c) Power spectral density of $(\Delta B_\parallel)^2$ at $\Bext \! = \! 0.2$~T obtained from traces such as in (a) (blue) and (b) (black). Transition from white spectrum at low frequencies to $1/f^2$ at high frequencies is reproduced by the nuclear spin diffusion model (gray). A deviation from this dependence at the highest frequencies is a numerical artifact caused by the discreteness of $|\Delta B_\parallel|$ values obtained from the Fourier analysis.
}
	\label{fig2}
\end{figure}

Dynamics of the S-T$_{0}$ qubit in the well-separated (1,1) charge state, i.e., for vanishing exchange, $J$, between the two electrons, is governed by the static external magnetic field $\Bext$ and dynamic Overhauser fields. For large $\Bext$, we can model the qubit evolution using the Hamiltonian~\cite{Malinowski2017,Bluhm2011,Neder2011}
\begin{equation}
	\label{H}
	\hat{H}(t) \! = \! g \mu_B \sum\limits_{i=L,R} \left( B_\parallel^i(t) \! + \! \frac{|\vec{B}_\perp^i(t)|^2}{2|\Bext|}\right) \hat{S}^i_z,
\end{equation}
where $g \sim -0.4$ is the electronic $g$-factor, $\mu_B$ is a Bohr magneton, $\hat{S}_z^i$ is the spin operator of the electron in left or right dot $i \! = \! L,R$, and $B_\parallel^i$ is the Overhauser field component parallel to $\Bext$. The influence of the transverse Overhauser field component $\vec{B}_\perp^i$ on the qubit is strongly suppressed when $\Bext$ is much larger than the typical Overhauser field. 
Hence the transverse Overhauser field fluctuations play a significant role in the qubit evolution only when the influence of the fluctuating longitudinal Overhauser field $B_\parallel^i$ is eliminated by dynamical decoupling~\cite{Bluhm2011, Malinowski2017}. The splitting between qubit states $\ket{\du}$ and $\ket{\ud}$ for $J=0$ is thus proportional to the longitudinal component of the Overhauser field gradient, $\Delta B_\parallel \! = \! B_\parallel^L \! - \! B_\parallel^R$, and can be measured by monitoring the qubit precession between $\ket{S}$ and $\ket{T_0}$~\cite{Foletti2010,Barthel2009,Barthel2012}. 

To measure this precession, we apply a cyclic pulse sequence that first prepares the singlet, then separates the two electrons to allow free precession in the Overhauser field for time $t_S$, and finally performs a projective readout of the qubit in the S-T$_{0}$ basis (Fig.~\ref{fig1}d). 
The total length of the pulse sequence is approximately 30 $\mu$s, including 10 $\mu$s of readout time. 
For each $t_S$ we use 16 single-shot readouts of this sequence to estimate the singlet return probability, $P_S$.
By repeatedly sweeping $t_S$ from 0 to 250~ns in 300 steps allows the precession of the qubit in the evolving Overhauser field to be measured with roughly 1~s temporal resolution (slow mode). 
A time trace showing 80~s  of slow-mode probability data is shown in Fig.~\ref{fig2}a. 
To increase the temporal resolution from 1 s to 12 ms we omit the probability estimation and record one single-shot outcome for each $t_S$ (fast mode). A time trace showing 1 s of fast-mode single-shot data is shown in Fig.~\ref{fig2}b. The time evolution of the qubit precession frequency, $f_\mathrm{Ovh}(t)$, is then extracted from these data as described in the Suppl. Section 1. The frequency corresponds to the absolute value of the Overhauser field gradient $|\Delta B_\parallel(t)| \! = \! h f_\mathrm{Ovh}(t)/|g| \mu_B$.
Examples of $|\Delta B_\parallel(t)|$ for $\Bext \! = \! 0.2$~T are shown in Figs.~\ref{fig2}a,b. In contrast to experiments performing dynamic nuclear polarization~\cite{Danon2009,Bluhm2010,Forster2015} the observed distributions of $\Delta B_\parallel$ reveal no sign of multistable behaviour (see  Suppl. Section~2).

\begin{figure}[tbh]
	\includegraphics[width=0.47\textwidth]{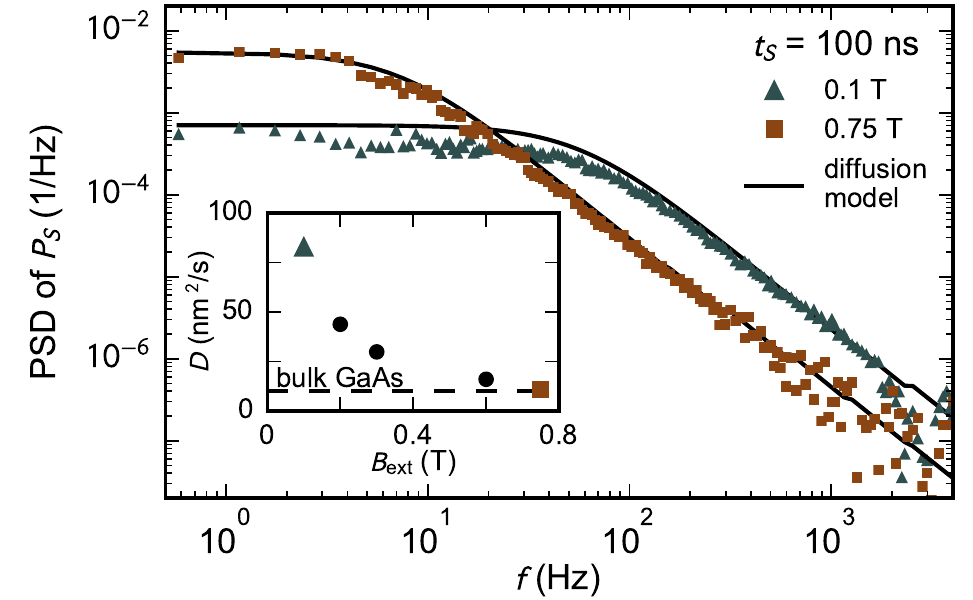}
	\caption{Magnetic field dependence of the power spectral density of $P_S$, keeping $t_S=100$~ns fixed. Increasing $\Bext$ from 0.1 to 0.75~T suppresses the $1/f^2$ noise by an order of magnitude. 
	Solid lines are fits of the diffusion model with the effective diffusion constant $D$ being the only free parameter.
	Inset: $D$ as a function of magnetic field $\Bext$. Dashed line indicates the spin diffusion constant for bulk GaAs, $D \! = \! 10$~$\mathrm{nm^2/s}$~\cite{Paget1982}.
	}
	\label{fig3}
\end{figure}

Next, we focus on the power spectral density (PSD) of $\Delta B_\parallel$ for $\Bext \! = \! 0.2$~T. Since taking the absolute value of $\Delta B_\parallel$ introduces kinks in $|\Delta B_\parallel|$ traces, adding spurious high-frequency content, we instead extract the PSD of $(\Delta B_\parallel)^2$ (Fig.~\ref{fig2}c). The resulting spectrum is flat below $10^{-2}$~Hz and falls off as $1/f^2$ above 1~Hz, indicating a correlation time of $\Delta B_\parallel$ of a few seconds.

A classical model of Overhauser field fluctuations due to nuclear spin diffusion is used to fit the experimental data in Fig.~\ref{fig2}c~\cite{Reilly2008} (Suppl. Section 5).
In the model we use the double dot geometry estimated from the lithographic dimensions of the device and the heterostructure growth parameters (distance between the dots $d \! = \! 150$~nm, dot diameter $\sigma_\perp \! = \! 40$~nm and width of the electron wave function in the crystal growth direction $\sigma_z \! = \! 7.5$~nm). We fit the effective diffusion constant $D \! = \! 33$~nm$^2$/s and the equilibrium width of the $\Delta B_\parallel$ distribution $\sigma_{\Delta B} \! = \! 6.0$~mT.
This model yields the power spectrum of $\Delta B_\parallel$, which has the same qualitative behavior as the spectrum of $(\Delta B_\parallel)^2$ -- it is flat at low frequencies ($<10^{-2}$~Hz) and falls off as $1/f^2$ at high frequencies ($> \! 1$~Hz).
Such a relation between the PSD of a Gaussian distributed variable and that of its square is expected whenever the PSD has a $1/f^\beta$ dependence over a wide frequency range~\cite{Cywinski2014}.

In order to extend the spectral range to higher frequencies we apply the pulse cycle with a fixed separation time $t_S \! = \! 100$~ns, acquiring a single-shot measurement every 30~$\mu$s. This can be visualized as a horizontal cut through the data in Fig.~\ref{fig2}b (top) at 100 ns, though, of course, now without taking the rest of the data at other values of $t_{S}$. Although the series of single-shot outcomes at fixed $t_{S}$ does not allow a direct measure of $\Delta B_\parallel$ from temporal oscillations, it does give statistical spectral information \cite{Reilly2008}. In particular, the Fourier transform of the windowed autocorrelation of single-shot outcomes (Suppl. Section 3) yields a PSD of the singlet return probability $P_S$, now extended to 4~kHz.

Power spectra of $P_{S}$ for the lowest and highest applied fields studied, $\Bext \! = \! 0.1$ and 0.75~T are shown in Fig.~\ref{fig3}. We observe that the spectrum for $\Bext \! = \! 0.75$~T is reduced by an order of magnitude in the $1/f^2$ regime, compared to the spectrum at $\Bext \! = \! 0.1$~T. To quantify the observed magnetic field dependence of the PSD of $P_S$ we fit the nuclear spin diffusion constant $D$ of the classical diffusion model~(Suppl. Section 5) to data, using fixed $\sigma_{\Delta B} \! = \! 6.0$~mT (obtained from the fit in Fig.~\ref{fig2}) and the same geometrical parameters as above.
The observed agreement with experimental data suggests that the effects of the nuclear spin bath are well described by classical evolution up to at least 1~kHz.

At low $\Bext$ we observe a strong enhancement of the effective spin diffusion constant compared to the literature value for bulk GaAs in the absence of free electrons, $D \! \sim \! 10$~$\mathrm{nm^2/s}$~\cite{Paget1982} (Fig.~\ref{fig3}, inset).
Qualitatively, this increase may be attributed to electron-mediated nuclear flip-flop processes~\cite{Klauser2008,Latta2011,Gong2011, Reilly2008,Reilly2010}, which dominate over nuclear dipole-dipole mediated diffusion.
At 0.75~T the effective diffusion constant drops down to the value for bulk GaAs. Despite this agreement, we note that our values for $D$ are not corrected for possible changes of electronic wavefunctions with increasing magnetic field. A quantitative statement about the underlying bare diffusion constant is difficult, as the fitting results for D are sensitive to assumptions about the spatial extent of the quantum dots (in particular $\sigma_\perp$) and the fraction of time spent in (1,1) and (2,0). 
Since spin diffusion due to nuclear dipole-dipole interaction is strongly suppressed by the Knight field gradient~\cite{Deng2005} and quadrupolar splittings, we expect further suppression of $D$ at higher magnetic fields~\cite{Gong2011}, and saturation below the bulk GaAs value. Indeed, this is observed in self-assembled quantum dots, where quadrupolar splittings are significantly stronger due to strain~\cite{Nikolaenko2009,Latta2011,Chekhovich2015}.

Overhauser field fluctuations above 100 kHz are too fast to be observed as oscillation between $\ket{S}$ and $\ket{T_0}$ with the present setup. However, we can infer spectral features from the decoherence of $\ket{\ud}$ and $\ket{\du}$ states using Hahn echo and Carr-Purcell-Meiboom-Gill (CPMG) dynamical decoupling sequences~\cite{Medford2012,Malinowski2017}. Since these decoupling sequences act as filters in frequency domain, we can relate the Overhauser spectrum to the decay of qubit coherence \cite{Malinowski2017,Martinis2003,Cywinski2008,Biercuk2010}. In particular, Hahn echo and CPMG sequences suppress the low frequency fluctuations, making the coherence decay a sensitive probe of high-frequency Overhauser fields. 

\begin{figure}[t!]
	\includegraphics[width=0.47\textwidth]{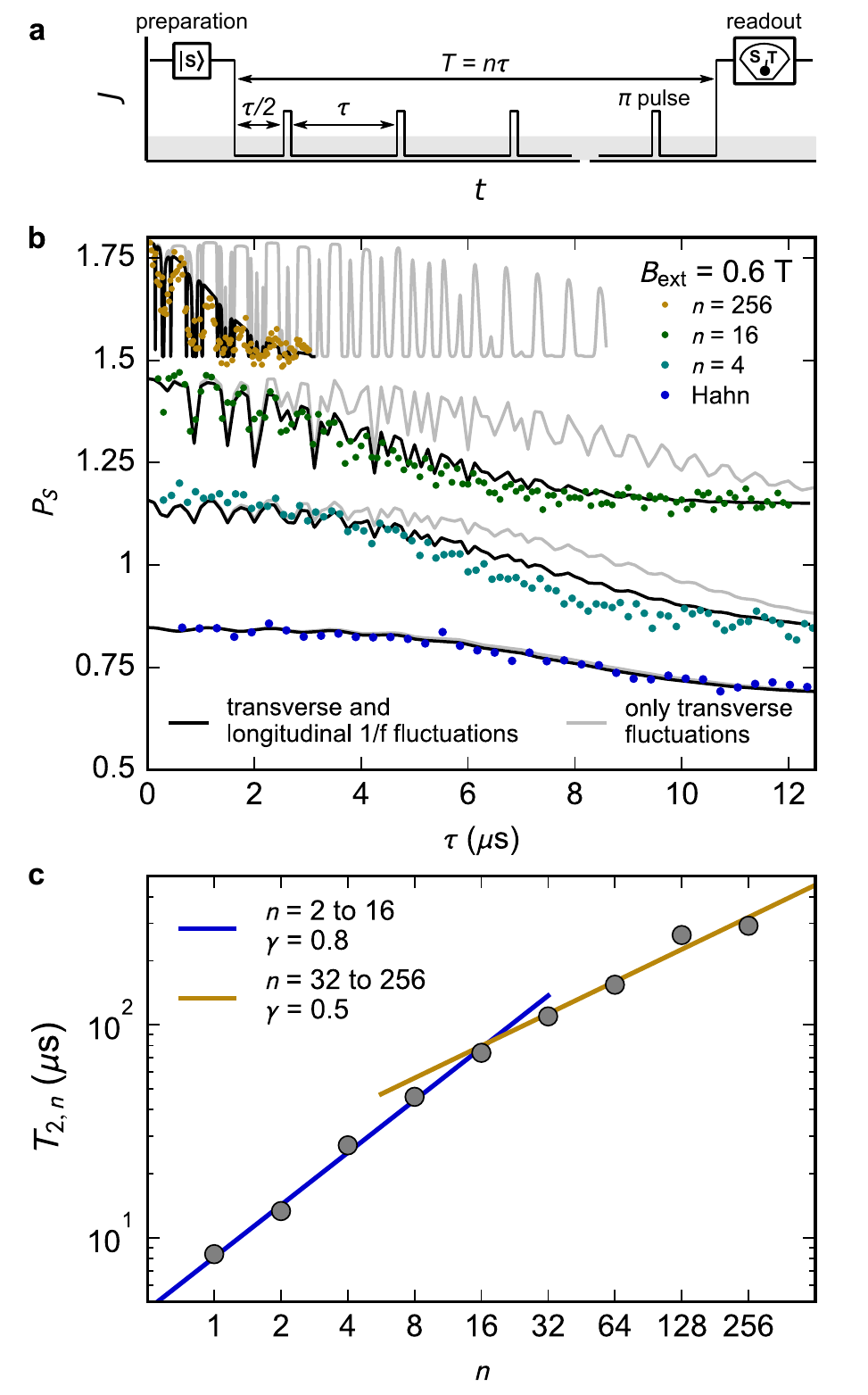}
	\caption{
	(a) Schematic of a CPMG dynamical decoupling sequence applied to a $S$-$T_0$ qubit, presented as a time dependent exchange energy $J$ (see text).
		(b) Coherence of the S-T$_0$ qubit after Hahn echo and CPMG sequences with number of $\pi$ pulses $n$. $\tau \! = \! T/n$ is the repetition period between pulses. Black curves present simulations including longitudinal $1/f$ noise and transverse fluctuations due to Larmor precession of the nuclei. Gray curves assume transverse Overhauser field fluctuations only. Data and curves are offset for clarity.
	(c) Scaling of the extracted coherence decay envelope $\Tn$ with $n$. Solid blue and yellow lines indicate fits of the power law $\propto \! n^\gamma$ to data in the indicated range.
	A large value of $\gamma \! = \! 0.8$ for small number of $\pi$ pulses indicates that decay is dominated by the transverse noise. $\gamma \! = \! 0.5$ for large $n$ is consistent with decay due to longitudinal $1/f$ noise.}
	\label{fig4}
\end{figure}

The decoupling sequence in Fig.~\ref{fig4}a uses symmetric exchange pulses \cite{Martins2016}, but is otherwise standard \cite{Medford2012}: initialize in S(2,0), evolve for time $\tau/2$ in (1,1), apply symmetric exchange $\pi$-pulse, evolve for another $\tau/2$, repeat the $\tau/2 \! - \! \pi \! - \! \tau/2$ segment a total of $n$ times. After the total evolution time $T \! = \! n\tau$, project onto S-T$_{0}$ by pulsing to (2,0) and perform single-shot readout.  Averaging $\sim$1000 such single-shot readouts then yields the singlet return probability. 
For such a sequence the resulting singlet return probability is related to the qubit coherence by $P_{S} = \frac{1}{2} + \frac{1}{2}\text{Re}[W_{L}(n\tau)W^{*}_{R}(n\tau) ]$, where $W_{i}(t)$ is the normalized coherence of the spin in dot $i$ at time $t$. 

Figure  \ref{fig4}b shows the singlet return probability for Hahn echo and CPMG sequences with various numbers of $\pi$ pulses, $n$, as a function of the interpulse time $\tau \! = \! T/n$.
For sequences with small $n$, coherence decreases smoothly with $\tau$, while for sequences with large $n$ the decay is strongly modulated. It was previously shown~\cite{Bluhm2011,Malinowski2017} that the coherence modulations are due to narrowband spectral content at megahertz frequencies in the transverse Overhauser field $\vec{B}^{i}_\perp$, arising from the relative Larmor precession of the three nuclear species.

The influence of  transverse Overhauser fluctuations, $\vec{B}^{i}_\perp$, on the CPMG signal decay was simulated using a semiclassical theory \cite{Cywinski2009,Cywinski2009a,Neder2011} that previously gave good agreement with echo \cite{Bluhm2010,Botzem2016} and CPMG \cite{Malinowski2017} experiments (see Suppl. Section 6 for details). 
Comparisons of experimental data with numerical simulations are shown in Fig.~\ref{fig4}b.
First, we include only narrowband transverse fields (gray curves), assuming two identical dots each containing $N = 9\times 10^5$ nuclei and a spread of effective fields experienced by the nuclei of $\delta B \! = \! 1$~mT, arising, for example, from quadrupolar splittings~\cite{Bluhm2011,Stockill2016,Botzem2016,Stockill2016}. This simulation reproduces the coherence decay for Hahn echo and the coherence modulations.
The decay envelopes for the simulated CPMG, however, do not agree well with experiment, especially for large $n$. In order to gain additional insight into the source of decoherence we extract the envelope decay time, $\Tn$, from the experimental data and plot it as a function of $n$ (Fig.~\ref{fig4}c and Suppl. Fig.~S4)~\cite{Medford2012}. 
We observe an initial scaling of $\TCPMG \! \propto \! n^\gamma$ with $\gamma \! \sim \! 0.8$, and a crossover to $\gamma \! \sim \! 0.5$ for large $n$.

We ascribe the change in the observed $\TCPMG$ scaling to a crossover between decoherence limited by transverse to longitudinal Overhauser field dynamics. For small $n$ the fluctuations of $\vec{B}^{i}_\perp$ dominate the decoherence, leading to scaling with large $\gamma$; purely transverse low-frequency fluctuations are expected to yield $\TCPMG \! \propto n^\gamma$ with $\gamma \! = \! 1$ (see Suppl. Section 6).
With increasing $n$ other decoherence sources start playing a dominant role. 
The intermediate-frequency fluctuations of $\Delta B_\parallel$ cause additional superexponential decay, which for large $n$ is given by $\exp[-4TS_{\parallel}(1/2\tau)/\pi^2]$, where $S_{\parallel}(f)$ is the PSD of $\Delta B_\parallel$~\cite{Alvarez2011,Yuge2011,Bylander2011}. 
Assuming that this PSD has a $1/f^{\beta}$ power-law behavior in the relevant frequency range, the CPMG decay for fixed $n$ and varying $\tau$ is then  $\exp[-(T/\Tn)^{\beta+1}]$, with $\Tn \! \propto \! n^{\gamma}$ and $\gamma \! = \! \beta/(\beta \! + \! 1)$~\cite{Medford2012}.
The observed scaling with $\gamma \! \sim \! 0.5$ is therefore consistent with $1/f$ noise and a Gaussian decay. 

As shown in Fig.~\ref{fig4}b (black lines), adding the $\beta = 1$ envelope function, $\exp[-(T/\Tn)^2]$ and $\Tn \! = \! n^{1/2} \! \times \! 25$~$\mu$s, appropriate for $\beta = 1$, gives good agreement with  experimental results. From the agreement between the simulations and the measurements we estimate that for $f \! > \! 100$~kHz the PSD $S_{\parallel}(f) \! \sim \! A^{2}/(2\pi f)$ with  $A^{-1} \! \sim \! 9$~$\mu$s.
For comparison with results presented in Ref.~\cite{Malinowski2017} we extrapolate this frequency dependence to $667$~kHz. Using the extrapolated value we estimate the CPMG decay time in an experiment in which $\tau$ is fixed but $n$ is varied, $\TCPMG \! = \! \pi^2/4S_{\parallel}(1/2\tau)$. Such estimate yields $\approx\!0.83$~ms for $\tau \! = \! 750$~ns, which is close to $\TCPMG \! = \! 0.87 \! \pm \! 0.13$~ms measured in Ref.~\cite{Malinowski2017}.

The $1/f$ power law found for $f \! > \! 100$~kHz differs from the $1/f^2$ spectrum observed below 1~kHz. This is not surprising, since for frequencies higher than the strength of intra-nuclear interactions ($\sim$1~kHz) the diffusion model is no longer applicable.
Whether the high-frequency $\Delta B_\parallel$ fluctuations have the same physical origin (i.e.~flip-flops of nuclei due to dipolar and hyperfine-mediated interactions) as the low-frequency ones is an open question.

Theory for CPMG decay caused by spectral diffusion due to dipolar interactions predicts a coherence decay of the form $\exp[-(T/\Tn)^6]$,  with $\Tn \! \propto \! n^{2/3}$ for small and even $n$ \cite{Witzel2007}.
This decay form (and scaling) is in disagreement with our observations.
In particular for large $n$, existing spectral diffusion theories based on cluster expansion~\cite{Witzel2006,Yao2006,Yang2009a} may need to be refined, for example taking into account realistic shapes of the electronic wave functions. Based on our findings, such theories can be tested experimentally at $\Bext>1$~T, where bare dipole-dipole coupling is the dominant internuclear interaction. 

Finally, it is possible that the $\Delta B_\parallel$ fluctuations are not of intrinsic origin (nuclear dynamics), but of extrinsic origin.
For example, charge noise, which generically has a $1/f^\beta$ spectrum with $\beta \! \sim \! 1$~\cite{Dial2013}, can shift the electron wavefunction and effectively result in Overhauser field fluctuations~\cite{Neder2011}.

In conclusion, we have experimentally investigated the spectrum of the GaAs nuclear environment for spin qubits and find it consistent with classical diffusion over six orders of magnitude in frequency, from millihertz to kilohertz. For applied fields below $\sim\!0.75$~T, nuclear diffusion is dominated by the electron-mediated flip-flop, enhancing diffusion by a factor of 8. Decoherence of the S-T$_0$ qubit is dominated by fluctuations of the transverse Overhauser field for short CPMG sequences, and by longitudinal Overhauser field for CPMG sequences with more than 32 $\pi$ pulses.

This work was supported by the Army Research Office, the Polish National Science Centre (NCN) under Grants No.~DEC-2012/07/B/ST3/03616 and DEC-2015/19/B/ST3/03152, the Innovation Fund Denmark, the Villum Foundation and the Danish National Research Foundation. 

\bibliographystyle{naturemag}
\bibliography{bibliography}{}

\begin{thebibliography}{10}
\expandafter\ifx\csname url\endcsname\relax
  \def\url#1{\texttt{#1}}\fi
\expandafter\ifx\csname urlprefix\endcsname\relax\def\urlprefix{URL }\fi
\providecommand{\bibinfo}[2]{#2}
\providecommand{\eprint}[2][]{\url{#2}}

\bibitem{Loss1998}

\newblock \bibinfo{author}{Loss, D.} \& \bibinfo{author}{DiVincenzo, D.~P.}
\newblock \emph{\bibinfo{journal}{Physical Review A}}
  \textbf{\bibinfo{volume}{57}}, \bibinfo{pages}{120--126}
  (\bibinfo{year}{1998}).

\bibitem{Veldhorst2015}

\newblock \bibinfo{author}{Veldhorst, M.} \emph{et~al.}
\newblock \emph{\bibinfo{journal}{Nature (London)}}
  \textbf{\bibinfo{volume}{526}}, \bibinfo{pages}{410--414}
  (\bibinfo{year}{2015}).

\bibitem{Petta2005}

\newblock \bibinfo{author}{Petta, J.} \emph{et~al.}
\newblock \emph{\bibinfo{journal}{Science}} \textbf{\bibinfo{volume}{309}},
  \bibinfo{pages}{2180--2184} (\bibinfo{year}{2005}).

\bibitem{Nowack2011}

\newblock \bibinfo{author}{Nowack, K.~C.} \emph{et~al.}
\newblock \emph{\bibinfo{journal}{Science}} \textbf{\bibinfo{volume}{333}},
  \bibinfo{pages}{1269--1272} (\bibinfo{year}{2011}).

\bibitem{Shulman2012}

\newblock \bibinfo{author}{Shulman, M.~D.}, \bibinfo{author}{Dial, O.~E.},
  \bibinfo{author}{Harvey, S.~P.}, \bibinfo{author}{Bluhm, H.},
  \bibinfo{author}{Umansky, V.} \& \bibinfo{author}{Yacoby, A.}
\newblock \emph{\bibinfo{journal}{Science}} \textbf{\bibinfo{volume}{336}},
  \bibinfo{pages}{202--205} (\bibinfo{year}{2012}).

\bibitem{Malinowski2017}

\newblock \bibinfo{author}{Malinowski, F.~K.} \emph{et~al.}
\newblock \emph{\bibinfo{journal}{Nature Nanotechnology}}
  \textbf{\bibinfo{volume}{12}}, \bibinfo{pages}{16--20}
  (\bibinfo{year}{2017}).

\bibitem{Bluhm2011}

\newblock \bibinfo{author}{Bluhm, H.}, \bibinfo{author}{Foletti, S.},
  \bibinfo{author}{Neder, I.}, \bibinfo{author}{Rudner, M.~S.},
  \bibinfo{author}{Mahalu, D.}, \bibinfo{author}{Umansky, V.} \&
  \bibinfo{author}{Yacoby, A.}
\newblock \emph{\bibinfo{journal}{Nature Physics}}
  \textbf{\bibinfo{volume}{7}}, \bibinfo{pages}{109--113}
  (\bibinfo{year}{2011}).

\bibitem{Foletti2010}

\newblock \bibinfo{author}{Foletti, S.}, \bibinfo{author}{Bluhm, H.},
  \bibinfo{author}{Mahalu, D.}, \bibinfo{author}{Umansky, V.} \&
  \bibinfo{author}{Yacoby, A.}
\newblock \emph{\bibinfo{journal}{Nature Physics}}
  \textbf{\bibinfo{volume}{5}}, \bibinfo{pages}{903--908}
  (\bibinfo{year}{2010}).

\bibitem{Barthel2009}

\newblock \bibinfo{author}{Barthel, C.}, \bibinfo{author}{Reilly, D.~J.},
  \bibinfo{author}{Marcus, C.~M.}, \bibinfo{author}{Hanson, M.~P.} \&
  \bibinfo{author}{Gossard, A.~C.}
\newblock \emph{\bibinfo{journal}{Physical Review Letters}}
  \textbf{\bibinfo{volume}{103}}, \bibinfo{pages}{160503}
  (\bibinfo{year}{2009}).

\bibitem{Shulman2014}

\newblock \bibinfo{author}{Shulman, M.~D.}, \bibinfo{author}{Harvey, S.~P.},
  \bibinfo{author}{Nichol, J.~M.}, \bibinfo{author}{Bartlett, S.~D.},
  \bibinfo{author}{Doherty, A.~C.}, \bibinfo{author}{Umansky, V.} \&
  \bibinfo{author}{Yacoby, A.}
\newblock \emph{\bibinfo{journal}{Nature Communications}}
  \textbf{\bibinfo{volume}{5}}, \bibinfo{pages}{5156} (\bibinfo{year}{2014}).

\bibitem{Bechtold2015}

\newblock \bibinfo{author}{Bechtold, A.} \emph{et~al.}
\newblock \emph{\bibinfo{journal}{Nature Physics}}
  \textbf{\bibinfo{volume}{11}}, \bibinfo{pages}{1005--1008}
  (\bibinfo{year}{2015}).

\bibitem{Nichol2015}

\newblock \bibinfo{author}{Nichol, J.~M.} \emph{et~al.}
\newblock \emph{\bibinfo{journal}{Nature communications}}
  \textbf{\bibinfo{volume}{6}}, \bibinfo{pages}{7682} (\bibinfo{year}{2015}).

\bibitem{Delbecq2016}

\newblock \bibinfo{author}{Delbecq, M.~R.} \emph{et~al.}
\newblock \emph{\bibinfo{journal}{Physical Review Letters}}
  \textbf{\bibinfo{volume}{116}}, \bibinfo{pages}{046802}
  (\bibinfo{year}{2016}).

\bibitem{Neder2011}

\newblock \bibinfo{author}{Neder, I.}, \bibinfo{author}{Rudner, M.~S.},
  \bibinfo{author}{Bluhm, H.}, \bibinfo{author}{Foletti, S.},
  \bibinfo{author}{Halperin, B.~I.} \& \bibinfo{author}{Yacoby, A.}
\newblock \emph{\bibinfo{journal}{Physical Review B}}
  \textbf{\bibinfo{volume}{84}}, \bibinfo{pages}{035441}
  (\bibinfo{year}{2011}).

\bibitem{Barthel2012}

\newblock \bibinfo{author}{Barthel, C.}, \bibinfo{author}{Medford, J.},
  \bibinfo{author}{Bluhm, H.}, \bibinfo{author}{Yacoby, A.},
  \bibinfo{author}{Marcus, C.~M.}, \bibinfo{author}{Hanson, M.~P.} \&
  \bibinfo{author}{Gossard, A.~C.}
\newblock \emph{\bibinfo{journal}{Physical Review B}}
  \textbf{\bibinfo{volume}{85}}, \bibinfo{pages}{035306}
  (\bibinfo{year}{2012}).

\bibitem{Danon2009}

\newblock \bibinfo{author}{Danon, J.}, \bibinfo{author}{Vink, I.~T.},
  \bibinfo{author}{Koppens, F. H.~L.}, \bibinfo{author}{Nowack, K.~C.},
  \bibinfo{author}{Vandersypen, L. M.~K.} \& \bibinfo{author}{Nazarov, Y.~V.}
\newblock \emph{\bibinfo{journal}{Physical Review Letters}}
  \textbf{\bibinfo{volume}{103}}, \bibinfo{pages}{046601}
  (\bibinfo{year}{2009}).

\bibitem{Bluhm2010}

\newblock \bibinfo{author}{Bluhm, H.}, \bibinfo{author}{Foletti, S.},
  \bibinfo{author}{Mahalu, D.}, \bibinfo{author}{Umansky, V.} \&
  \bibinfo{author}{Yacoby, A.}
\newblock \emph{\bibinfo{journal}{Physical Review Letters}}
  \textbf{\bibinfo{volume}{105}}, \bibinfo{pages}{216803}
  (\bibinfo{year}{2010}).

\bibitem{Forster2015}

\newblock \bibinfo{author}{Forster, F.}, \bibinfo{author}{Muhlbacher, M.},
  \bibinfo{author}{Schuh, D.}, \bibinfo{author}{Wegscheider, W.},
  \bibinfo{author}{Giedke, G.} \& \bibinfo{author}{Ludwig, S.}
\newblock \emph{\bibinfo{journal}{Physical Review B}}
  \textbf{\bibinfo{volume}{92}}, \bibinfo{pages}{6--10} (\bibinfo{year}{2015}).

\bibitem{Paget1982}

\newblock \bibinfo{author}{Paget, D.}
\newblock \emph{\bibinfo{journal}{Physical Review B}}
  \textbf{\bibinfo{volume}{25}}, \bibinfo{pages}{4444--4451}
  (\bibinfo{year}{1982}).

\bibitem{Reilly2008}

\newblock \bibinfo{author}{Reilly, D.~J.}, \bibinfo{author}{Taylor, J.},
  \bibinfo{author}{Laird, E.~A.}, \bibinfo{author}{Petta, J.},
  \bibinfo{author}{Marcus, C.~M.}, \bibinfo{author}{Hanson, M.~P.} \&
  \bibinfo{author}{Gossard, A.~C.}
\newblock \emph{\bibinfo{journal}{Physical Review Letters}}
  \textbf{\bibinfo{volume}{101}}, \bibinfo{pages}{236803}
  (\bibinfo{year}{2008}).

\bibitem{Cywinski2014}

\newblock \bibinfo{author}{Cywinski, L.}
\newblock \emph{\bibinfo{journal}{Physical Review A}}
  \textbf{\bibinfo{volume}{90}}, \bibinfo{pages}{042307}
  (\bibinfo{year}{2014}).

\bibitem{Klauser2008}

\newblock \bibinfo{author}{Klauser, D.}, \bibinfo{author}{Coish, W.~A.} \&
  \bibinfo{author}{Loss, D.}
\newblock \emph{\bibinfo{journal}{Physical Review B}}
  \textbf{\bibinfo{volume}{78}}, \bibinfo{pages}{205301}
  (\bibinfo{year}{2008}).

\bibitem{Latta2011}

\newblock \bibinfo{author}{Latta, C.}, \bibinfo{author}{Srivastava, A.} \&
  \bibinfo{author}{Imamoglu, A.}
\newblock \emph{\bibinfo{journal}{Physical Review Letters}}
  \textbf{\bibinfo{volume}{107}}, \bibinfo{pages}{167401}
  (\bibinfo{year}{2011}).

\bibitem{Gong2011}

\newblock \bibinfo{author}{Gong, Z.~X.}, \bibinfo{author}{Yin, Z.~Q.} \&
  \bibinfo{author}{Duan, L.~M.}
\newblock \emph{\bibinfo{journal}{New Journal of Physics}}
  \textbf{\bibinfo{volume}{13}}, \bibinfo{pages}{033036}
  (\bibinfo{year}{2011}).

\bibitem{Reilly2010}

\newblock \bibinfo{author}{Reilly, D.~J.}, \bibinfo{author}{Taylor, J.},
  \bibinfo{author}{Petta, J.}, \bibinfo{author}{Marcus, C.~M.},
  \bibinfo{author}{Hanson, M.~P.} \& \bibinfo{author}{Gossard, A.~C.}
\newblock \emph{\bibinfo{journal}{Physical Review Letters}}
  \textbf{\bibinfo{volume}{104}}, \bibinfo{pages}{236802}
  (\bibinfo{year}{2010}).

\bibitem{Deng2005}

\newblock \bibinfo{author}{Deng, C.} \& \bibinfo{author}{Hu, X.}
\newblock \emph{\bibinfo{journal}{Physical Review B}}
  \textbf{\bibinfo{volume}{72}}, \bibinfo{pages}{165333}
  (\bibinfo{year}{2005}).

\bibitem{Nikolaenko2009}

\newblock \bibinfo{author}{Nikolaenko, A.~E.} \emph{et~al.}
\newblock \emph{\bibinfo{journal}{Physical Review B}}
  \textbf{\bibinfo{volume}{79}}, \bibinfo{pages}{081303(R)}
  (\bibinfo{year}{2009}).

\bibitem{Chekhovich2015}

\newblock \bibinfo{author}{Chekhovich, E.~A.}, \bibinfo{author}{Hopkinson, M.},
  \bibinfo{author}{Skolnick, M.~S.} \& \bibinfo{author}{Tartakovskii, A.~I.}
\newblock \emph{\bibinfo{journal}{Nature Communications}}
  \textbf{\bibinfo{volume}{6}}, \bibinfo{pages}{6348} (\bibinfo{year}{2015}).

\bibitem{Medford2012}

\newblock \bibinfo{author}{Medford, J.}, \bibinfo{author}{Cywinski, L.},
  \bibinfo{author}{Barthel, C.}, \bibinfo{author}{Marcus, C.~M.},
  \bibinfo{author}{Hanson, M.~P.} \& \bibinfo{author}{Gossard, A.~C.}
\newblock \emph{\bibinfo{journal}{Physical Review Letters}}
  \textbf{\bibinfo{volume}{108}}, \bibinfo{pages}{086802}
  (\bibinfo{year}{2012}).

\bibitem{Martinis2003}

\newblock \bibinfo{author}{Martinis, J.~M.}, \bibinfo{author}{Nam, S.},
  \bibinfo{author}{Aumentado, J.}, \bibinfo{author}{Lang, K.} \&
  \bibinfo{author}{Urbina, C.}
\newblock \emph{\bibinfo{journal}{Physical Review B}}
  \textbf{\bibinfo{volume}{67}}, \bibinfo{pages}{94510} (\bibinfo{year}{2003}).

\bibitem{Cywinski2008}

\newblock \bibinfo{author}{Cywinski, L.}, \bibinfo{author}{Lutchyn, R.~M.},
  \bibinfo{author}{Nave, C.~P.} \& \bibinfo{author}{{Das Sarma}, S.}
\newblock \emph{\bibinfo{journal}{Physical Review B}}
  \textbf{\bibinfo{volume}{77}}, \bibinfo{pages}{174509}
  (\bibinfo{year}{2008}).

\bibitem{Biercuk2010}

\newblock \bibinfo{author}{Biercuk, M.~J.}, \bibinfo{author}{Doherty, A.~C.} \&
  \bibinfo{author}{Uys, H.}
\newblock \emph{\bibinfo{journal}{Journal of Physics B: Atomic, Molecular and
  Optical Physics}} \textbf{\bibinfo{volume}{44}}, \bibinfo{pages}{154002}
  (\bibinfo{year}{2010}).

\bibitem{Martins2016}

\newblock \bibinfo{author}{Martins, F.} \emph{et~al.}
\newblock \emph{\bibinfo{journal}{Physical Review Letters}}
  \textbf{\bibinfo{volume}{116}}, \bibinfo{pages}{116801}
  (\bibinfo{year}{2016}).

\bibitem{Cywinski2009}

\newblock \bibinfo{author}{Cywinski, L.}, \bibinfo{author}{Witzel, W.~M.} \&
  \bibinfo{author}{{Das Sarma}, S.}
\newblock \emph{\bibinfo{journal}{Physical Review Letters}}
  \textbf{\bibinfo{volume}{102}}, \bibinfo{pages}{057601}
  (\bibinfo{year}{2009}).

\bibitem{Cywinski2009a}

\newblock \bibinfo{author}{Cywinski, L.}, \bibinfo{author}{Witzel, W.~M.} \&
  \bibinfo{author}{{Das Sarma}, S.}
\newblock \emph{\bibinfo{journal}{Physical Review B}}
  \textbf{\bibinfo{volume}{79}}, \bibinfo{pages}{245314}
  (\bibinfo{year}{2009}).

\bibitem{Botzem2016}

\newblock \bibinfo{author}{Botzem, T.}, \bibinfo{author}{McNeil, R. P.~G.},
  \bibinfo{author}{Schuh, D.}, \bibinfo{author}{Bougeard, D.} \&
  \bibinfo{author}{Bluhm, H.}
\newblock \emph{\bibinfo{journal}{Nature communications}}
  \textbf{\bibinfo{volume}{7}}, \bibinfo{pages}{11170} (\bibinfo{year}{2016}).

\bibitem{Stockill2016}

\newblock \bibinfo{author}{Stockill, R.}, \bibinfo{author}{{Le Gall}, C.},
  \bibinfo{author}{Matthiesen, C.}, \bibinfo{author}{Huthmacher, L.},
  \bibinfo{author}{Clarke, E.}, \bibinfo{author}{Hugues, M.} \&
  \bibinfo{author}{Atat{\"{u}}re, M.}
\newblock \emph{\bibinfo{journal}{Nature Communications}}
  \textbf{\bibinfo{volume}{7}}, \bibinfo{pages}{12745} (\bibinfo{year}{2016}).

\bibitem{Alvarez2011}

\newblock \bibinfo{author}{{\'{A}}lvarez, G.~A.} \& \bibinfo{author}{Suter, D.}
\newblock \emph{\bibinfo{journal}{Physical Review Letters}}
  \textbf{\bibinfo{volume}{107}}, \bibinfo{pages}{230501}
  (\bibinfo{year}{2011}).

\bibitem{Yuge2011}

\newblock \bibinfo{author}{Yuge, T.}, \bibinfo{author}{Sasaki, S.} \&
  \bibinfo{author}{Hirayama, Y.}
\newblock \emph{\bibinfo{journal}{Physical Review Letters}}
  \textbf{\bibinfo{volume}{107}}, \bibinfo{pages}{170504}
  (\bibinfo{year}{2011}).

\bibitem{Bylander2011}

\newblock \bibinfo{author}{Bylander, J.} \emph{et~al.}
\newblock \emph{\bibinfo{journal}{Nature Physics}}
  \textbf{\bibinfo{volume}{7}}, \bibinfo{pages}{565--570}
  (\bibinfo{year}{2011}).

\bibitem{Witzel2007}

\newblock \bibinfo{author}{Witzel, W.~M.} \& \bibinfo{author}{{Das Sarma}, S.}
\newblock \emph{\bibinfo{journal}{Physical Review Letters}}
  \textbf{\bibinfo{volume}{98}}, \bibinfo{pages}{077601}
  (\bibinfo{year}{2007}).

\bibitem{Witzel2006}

\newblock \bibinfo{author}{Witzel, W.~M.} \& \bibinfo{author}{{Das Sarma}, S.}
\newblock \emph{\bibinfo{journal}{Physical Review B}}
  \textbf{\bibinfo{volume}{74}}, \bibinfo{pages}{035322}
  (\bibinfo{year}{2006}).

\bibitem{Yao2006}

\newblock \bibinfo{author}{Yao, W.}, \bibinfo{author}{Liu, R.~B.} \&
  \bibinfo{author}{Sham, L.~J.}
\newblock \emph{\bibinfo{journal}{Physical Review B}}
  \textbf{\bibinfo{volume}{74}}, \bibinfo{pages}{195301}
  (\bibinfo{year}{2006}).

\bibitem{Yang2009a}

\newblock \bibinfo{author}{Yang, W.} \& \bibinfo{author}{Liu, R.~B.}
\newblock \emph{\bibinfo{journal}{Physical Review B}}
  \textbf{\bibinfo{volume}{79}}, \bibinfo{pages}{115320}
  (\bibinfo{year}{2009}).

\bibitem{Dial2013}

\newblock \bibinfo{author}{Dial, O.~E.}, \bibinfo{author}{Shulman, M.~D.},
  \bibinfo{author}{Harvey, S.~P.}, \bibinfo{author}{Bluhm, H.},
  \bibinfo{author}{Umansky, V.} \& \bibinfo{author}{Yacoby, A.}
\newblock \emph{\bibinfo{journal}{Physical Review Letters}}
  \textbf{\bibinfo{volume}{110}}, \bibinfo{pages}{146804}
  (\bibinfo{year}{2013}).

\end{thebibliography}

\onecolumngrid

\newpage

\renewcommand{\thefigure}{S\arabic{figure}}  
\renewcommand{\theequation}{S\arabic{equation}}
\renewcommand{\thetable}{S\arabic{table}}
\setcounter{figure}{0}   
\setcounter{equation}{0}   

\renewcommand{\theHfigure}{S.\thefigure}

\begin{center}
	\large \bf Spectrum of the Nuclear Environment for GaAs Spin Qubits -- supplementary information
\end{center}

\begin{center}

Filip K. Malinowski, Frederico Martins, {\L}ukasz Cywi\'nski, Mark S. Rudner, Peter D. Nissen, Saeed Fallahi, Geoffrey~C.~Gardner, Michael J. Manfra, Charles M. Marcus, and Ferdinand Kuemmeth
\vspace{0.5cm}

\begin{minipage}{0.8\textwidth}
	\small
	This supplementary information discusses the following topics:
		\begin{enumerate}
		\item Extracting the frequency of oscillations in Overhauser field from averaged and single-shot data
		\item Gaussian distribution of $\Delta B_\parallel$
		\item Obtaining power spectral density of $P_S$ from truncated autocorrelation of single-shot measurements
		\item Fitting procedures for PSD in Figs. 2c and 3
		\item Classical model of Overhauser field noise due to nuclear spin diffusion
		\item Decoherence of the qubit subjected to the transverse Overhauser noise
	\end{enumerate}
\end{minipage}
\end{center}

\twocolumngrid

\section{1. Extracting the frequency of oscillations in Overhauser field from averaged and single-shot data}

\begin{figure*}
	\includegraphics[scale=.9]{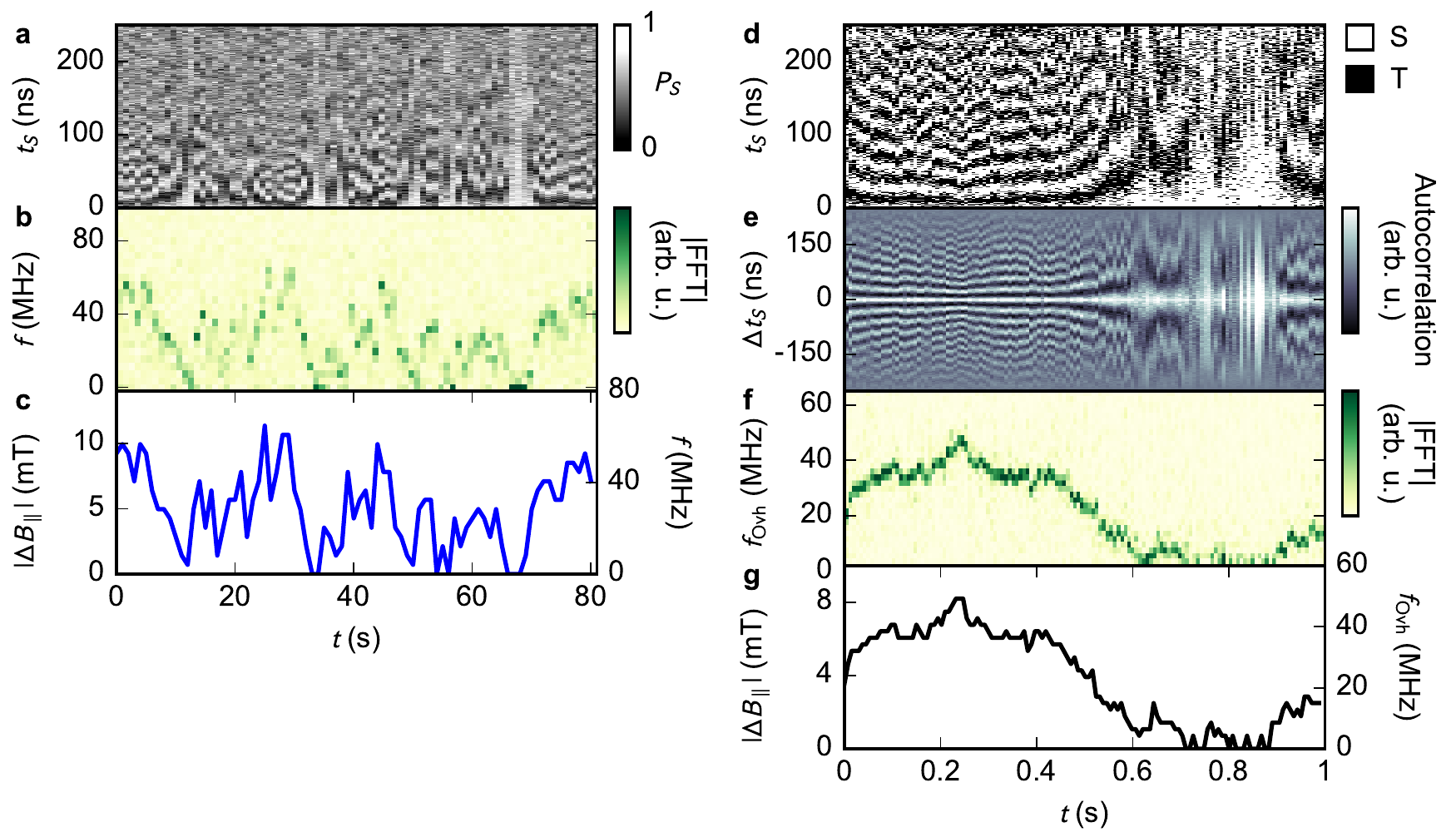}
	\caption{Intermediate steps of extracting $\Delta B_\parallel$ from raw data.
	(a) Raw data is taken by repeating a 300-point sequence with $t_S$ from 0 to 250~ns 16 times, allowing us to estimate the probability $P_S(t_S)$. Such data forms a single column of the presented map. (b) Absolute value of FFT of probability oscillations. (c) Extracted position of the peak, $f$, and corresponding $|\Delta B_\parallel|$.
	(d) Raw single-shot data (one sequence run per column) and (e) its autocorrelation. Color scale in autocorrelation 2d map is chosen to show oscillations and hide the peak at $\Delta t_S = 0$. (f) Absolute value of FFT of autocorrelation. (g) Extracted position of the peak and corresponding $|\Delta B_\parallel|$.}
	\label{figS1}
\end{figure*}

The power spectral density of the gradient of Overhauser field squared $(\Delta B_\parallel)^2$ is obtained from two kinds of data sets. 

The first one consists of oscillations in the singlet probability $P_S$ as a function of electron separation time $t_S$, varying from 0 to 250~ns, measured with 1~s repetition rate. 
For each studied value of magnetic field, we measure a 1.5~hour-long data set, and a fragment of such a data set is presented in Fig.~\ref{figS1}a. The frequencies of the oscillations are obtained by means of Fourier analysis. We calculate the Fast Fourier Transform (FFT) of each vertical column and inspect its absolute value (Fig.~\ref{figS1}b). The position of the maximum indicates the frequency of oscillations, which is related to the gradient of the Overhauser field between the dots by $hf_\mathrm{Ovh} = g\mu |\Delta B_\parallel|$ (Fig.~\ref{figS1}c).

The second data set consists of ten non-averaged measurements, each 30~s long.
A 1~s excerpt of one of them is shown in Fig.~\ref{figS1}d. 
Extracting the underlying oscillation frequency from each column requires more careful treatment, since the probabilistic nature of binary measurements adds large amounts of shot noise. 
In our analysis we first assign S and T$_0$ outcomes to, respectively, $1$ and $-1$. 
Then we subtract from each column its mean, and calculate the autocorrelation (Fig.~\ref{figS1}e).  
The obtained autocorrelation reveals oscillations at the same frequency as unprocessed data. However, in the autocorrelated data the shot noise is averaged out for all $\Delta t_S$ except for 0, where the shot noise accumulates. Next we replace the autocorrelation value at $\Delta t_S = 0$ with the value at the smallest $|\Delta t_S|\neq 0$. This minimizes the influence of shot noise without affecting the visibility of the oscillations. The absolute value of the FFT (Fig.~\ref{figS1}f) of the autocorrelations processed in such way exhibits a clear peak, which we associate with the qubit oscillation frequency.  
Namely, the peak position in frequency, $f_\mathrm{Ovh}$, is used to extract $|\Delta B_\parallel|$ via $hf_\mathrm{Ovh} = g\mu |\Delta B_\parallel|$ (Fig.~\ref{figS1}g).

\section{2. Gaussian distribution of $\Delta B_\parallel$}

\begin{figure}
	\includegraphics[scale=.9]{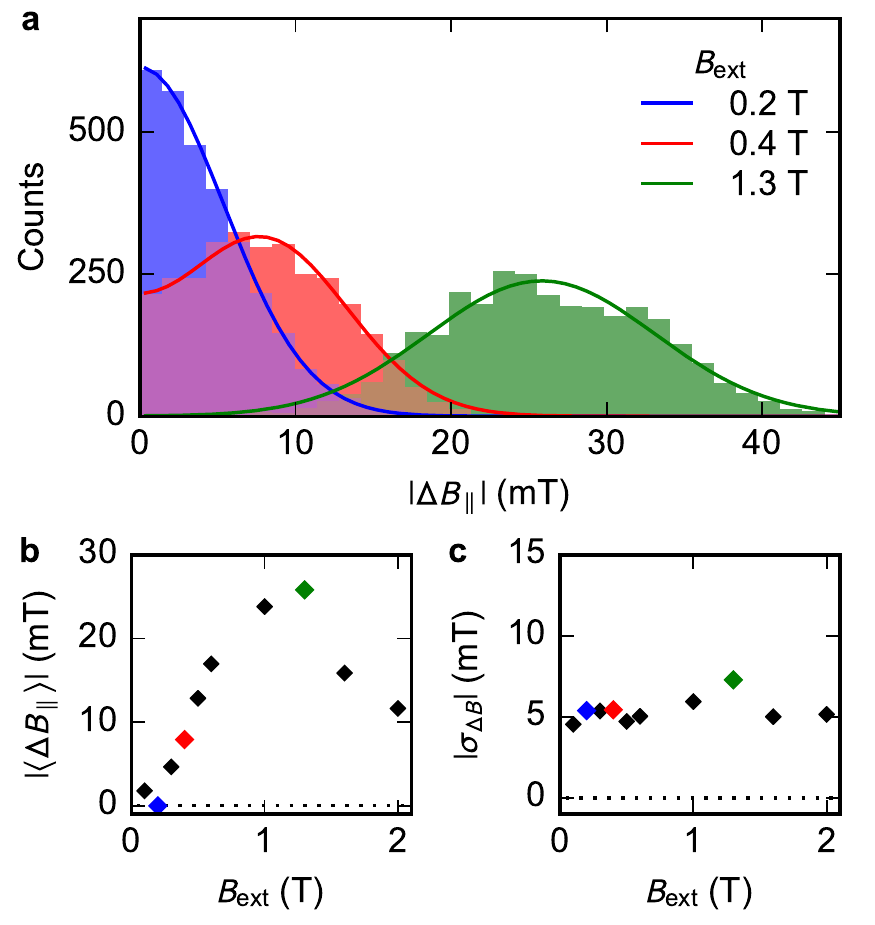}
	\caption{
	(a) Measured distribution of $|\Delta B_\parallel|$ for various applied magnetic fields $\Bext$. Solid lines are fits assuming Gaussian distributed $\Delta B_\parallel$ with mean $|\langle \Delta B_\parallel\rangle |$.
	(b,c) Fitted mean Overhauser field gradient $|\langle \Delta B_\parallel\rangle |$ and the Overhauser field gradient distribution width $\sigma_{\Delta B}$ as a function of applied magnetic field $\Bext$. Colored points correspond to $|\Delta B_\parallel|$ distributions presented in panel (a).
	}
	\label{figS2}
\end{figure}

Several points of the analysis presented in the main text assume a Gaussian distribution of the Overhauser field gradients. To show that this assumption is justified we plot in Fig.~\ref{figS2}a histograms of $|\Delta B_\parallel|$ for several values of the external magnetic field. The fits to the data confirm our assumption, indicating there spin bath does not have multiple stable points, in contract to experiments involving intentional dynamical nuclear polarization~\cite{Danon2009,Bluhm2010,Forster2015}. However we observe that the mean Overhauser field gradient is shfted away from zero for increasing external magnetic field, while the width of the $\Delta B_\parallel$ distribution remains unchanged (Fig.~\ref{figS2}b,c). We suspect that this non-zero mean arises from unintentional nuclear polarization, as reported previously for this device~\cite{Martins2016,Malinowski2017} and for devices studied by other groups~\cite{Botzem2016}.

\section{3. Obtaining power spectral density of $P_S$ from truncated autocorrelation of single-shot measurements}

\begin{figure*}
	\includegraphics[scale=.9]{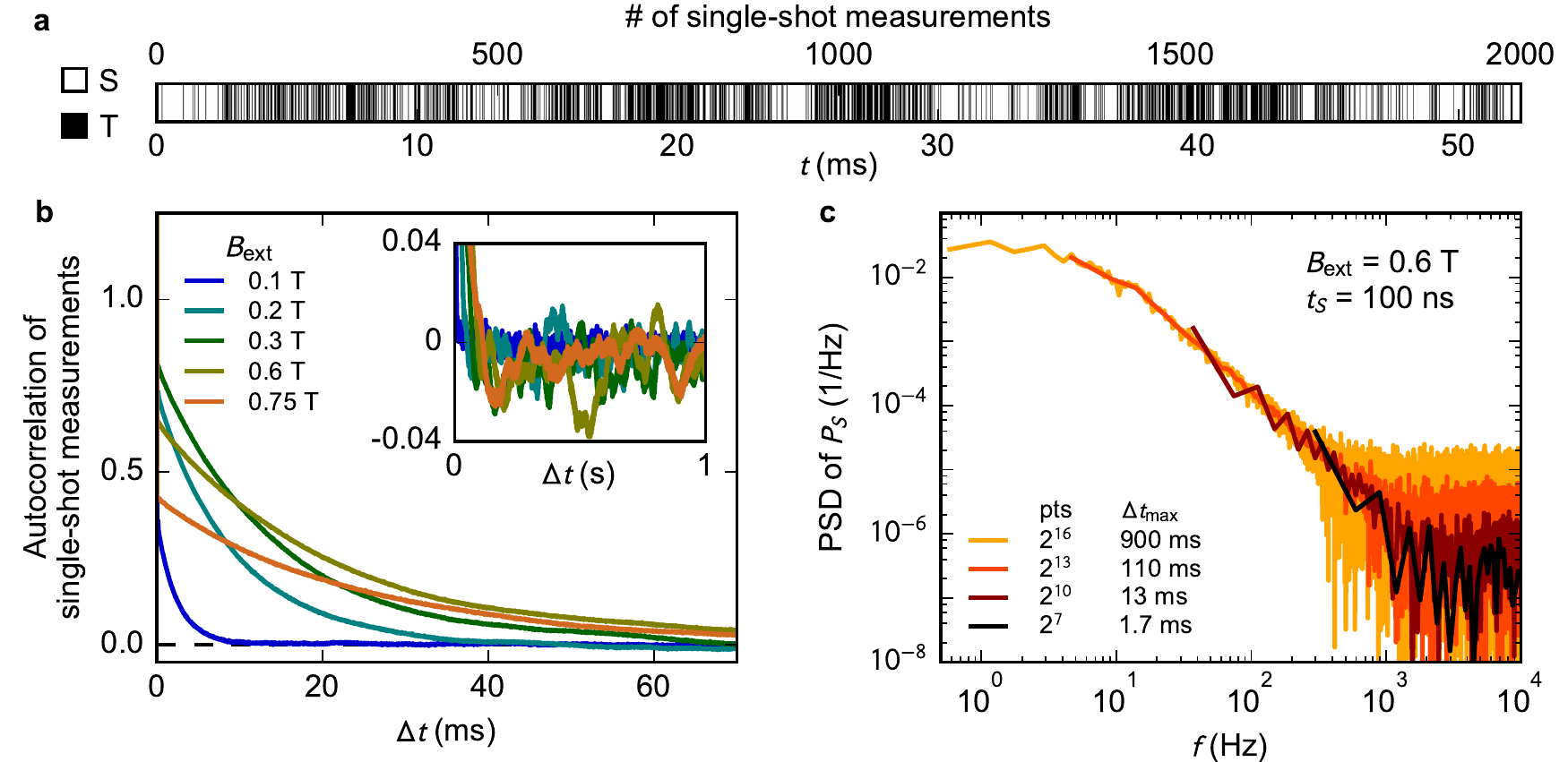}
	\caption{(a) Example trace of single-shot measurements obtained from pulses with $t_S=100$~ns at 0.6~T.
	(b) Autocorrelation of single-shot traces for $t_S=100$~ns. Inset shows long-time tail with oscillations caused by finite size of the sample.
	(c) Power spectral density of $P_S$ obtained by FFT of the autocorrelation truncated at $\pm \Delta t_\mathrm{max}$.}
	\label{figS3}
\end{figure*}

As explained in the main text, to maximize the repetition rate at which $\Delta B_\parallel$ is probed, we fix the separation time $t_S=100$~ns and repeat the pulse cycle continuously. Then we map S and T$_0$ outcomes to, respectively, $1$ and $-1$. As a result we obtain binary traces of over 2 million points. A small piece of such a trace is presented in Fig.~\ref{figS3}a, obtained for external magnetic field $\Bext = 0.6$~T and $t_S=100$~ns.

The sequence of single-shot outcomes is dominated by shot noise, which obscures the underlying oscillations when using conventional methods for calculating the PSD. To eliminate this noise contribution, we apply the same procedure mentioned in the previous section. That is, we find the autocorrelation and replace its value at $\Delta t=0$ with the value at the smallest $|\Delta t|\neq0$. The autocorrelation of single-shot measurements for $\Bext = 0.6$~T are plotted in Fig.~\ref{figS3}b. Now we can take advantage of the fact that the Fourier transform of the autocorrelation is identical to the power spectral density of the original trace.

Even though the sample is huge, we observe artifacts related to its finite size. Namely, the autocorrelation has a long, irregular tail (Fig.~\ref{figS3}b, inset). If we perform FFT over the entire available range of $\Delta t$, fluctuations in the tail dominate over the relevant features at $\Delta t \sim 1/f$.

In our further analysis we assume that relevant information about the nuclear noise at frequency $f$ is contained within the window $|\Delta t| \lesssim A/f = \Delta t_\mathrm{max}$ where $A$ is of the order of ten. In other words, to obtain an accurate value of the power spectral density of the noise at frequency $f$, it is sufficient to take the Fourier transform of the autocorrelation in the range $-\Delta t_\mathrm{max}<\Delta t<\Delta t_\mathrm{max}$. In our analysis we use $16\leq A\leq32$. 
To avoid the necessity of windowing we keep the range of $\Delta t$ such that the number of points within the $-\Delta t_\mathrm{max}<\Delta t<\Delta t_\mathrm{max}$ range is $2^n$, for integer $n$.

Power spectral densities for $\Bext = 0.6$~T, $t_S=100$~ns, and various choices of $\Delta t_\mathrm{max}$ are presented in Fig.~\ref{figS3}c, visualizing the trade-off between frequency range and noise floor level due to finite sample size. 
Wide $\Delta t$ windows (i.e. large $t_\mathrm{max}$) give access to lower frequencies but raise the noise floor, while narrow $\Delta t$ windows sacrifice low-frequency information for a reduction in noise. By adjusting the window dynamically we are able to achieve a wide spectral range without suffering from the background noise.

\section{4. Fitting procedures for PSD in~Figs.~2c~and~3}

The classical diffusion model used to describe the experimental results presented in Figs.~2c~and~3, provides analytical expressions for the autocorrelation of $(\Delta B_\parallel)^2$ and $P_S$, but no analytical expression for the PSD. Therefore the fitted expressions involve numerical Fourier transforms of the autocorrelation obtained from the analytical formulas.

To fit data in Fig.~2c we simulate two sets of autocorrelation traces, such that after performing a FFT they produce PSD points at the same frequencies as in the experimental data. Simulated traces obtained in such way are suitable for optimization via the method of least squares.

To fit the data in Fig.~3 we simulate autocorrelation traces with identical time resolution as the experimental data. To these traces we then apply the same procedure as to experimental data (that is we perform FFT of data limited to a suitable range of $\Delta t$). The obtained set of frequencies is identical to those of the experimental PSD, making this method suitable for least squares fitting.

\section{5. Classical model of Overhauser field noise due to nuclear spin diffusion}

Here we construct a model for the dynamics of a $S$-$T_0$ qubit in a double quantum dot, arising from slow fluctuations of the longitudinal Overhauser (difference) field in the two dots.
The Overhauser field is produced by nuclear spins in the host crystal, which undergo their own dynamics due to their mutual dipole-dipole coupling.
These dynamics lead to fluctuations of the longitudinal Overhauser field, which in turn affect the evolution of the spin qubit, which plays the role of a sensor in this work. 
Our approach is similar to that employed previously to describe the results of the experiment in Ref.~\cite{Reilly2008}.

The electronic system is influenced by the nuclear spins through the hyperfine (hf) interaction,
\begin{equation}
H_{\rm HF} = A_0 \sum_j \delta(\hat{\vec{r}} - \vec{R}_j)\, \hat{\vec{S}} \cdot \hat{\vec{I}}_j,
\end{equation}
where $\hat{\vec{r}}$ is the electron position operator, $\vec{R}_j$ is the position of nucleus $j$, and $\hat{\vec{S}}$ and $\hat{\vec{I}}_j$ are the spin operators for the electron and nucleus $j$, respectively.
Here $A_0$ has units $[{\rm Energy\ }\times\ {\rm Volume}/\hbar^2]$, with a characteristic value in GaAs of $\hbar^2A_0/v_0 \approx 100\ \mu {\rm eV}$ \cite{Cywinski2009a}, where $v_0$ is the unit cell volume.
For simplicity we consider a single nuclear species.
On a coarse-grained scale encompassing many atomic sites, we describe the nuclear spin state in terms of a spin density field $I(\vec{x},t)$.
Throughout this treatment we focus only on the spin component parallel to the externally-applied magnetic field.
Here $I(\vec{x},t)$ has units $[\hbar/{\rm Volume}]$.

In a double dot, the $(1,1)$ singlet and triplet ($T_0$) states are coupled by the longitudinal Overhauser difference field
\begin{equation}
  \Delta B_z(t) = \frac{\hbar A_0}{g_*\mu_{\rm B}} \int d^3x\, \Delta\rho(\vec{x})  I(\vec{x},t),  
\end{equation}
where $g_*$ is the electronic effective $g$-factor, $\mu_{\rm B}$ is the Bohr magneton, and $\Delta\rho(\vec{x}) =|\psi_R(\vec{x})|^2 - |\psi_L(\vec{x})|^2$, with $|\psi_R(\vec{x})|^2$ and $|\psi_L(\vec{x})|^2$ the electronic density profiles in the right and left dot, respectively.
Later it will be convenient to work in Fourier space,
\begin{equation}
 \label{eq:dBzFourier} \Delta B_z(t) = \frac{\hbar A_0}{g_*\mu_{\rm B}} \int \frac{d^3q}{(2\pi)^3}\, \Delta\tilde\rho_{\vec{q}}\,  \tilde I_{-\vec{q}}(t),
\end{equation}
where $\tilde{f}_{\vec{q}} = \int d^3x\, e^{-i\vec{q}\cdot\vec{x}} f(\vec{x})$.
Notably, although the nuclear spin field $I(\vec{x},t)$ extends throughout the entire sample, the Overhauser field is only sensitive to the value of $I(\vec{x},t)$ in a limited region where the electrons are localized.

For simplicity, we take a model where the nuclear spin field $I(\vec{x},t)$ evolves under its own dynamics, unperturbed by the presence of the electronic system.
In the absence of nuclear spin relaxation (i.e., for infinite nuclear $T_1$), the dipolar interaction between nuclear spins leads to a diffusive-type dynamics of nuclear spin polarization:
\begin{equation}
\label{eq:Diff_realspace}\partial_t I(\vec{x},t) = D \nabla^2 I(\vec{x},t) + \xi(\vec{x},t),
\end{equation}
where $\xi(\vec{x},t)$ is a stochastic field 
that accounts for the randomness of dipole-dipole induced nuclear spin flips.
The units of $\xi(\vec{x},t)$ are [Energy / Volume].
Such a diffusive model is also expected to at least qualitatively describe the dynamics caused by electron-mediated nuclear flip-flops~\cite{Gong2011}.

The smooth diffusive dynamics, as described by the first term in Eq.~(\ref{eq:Diff_realspace}), are only manifested on timescales longer than that for a single nuclear spin flip due to its interaction with its neighbors (typically $\sim 10-100\ \mu$s for GaAs \cite{Cywinski2009a}).  
On times longer than this scale, where the diffusion model applies, the noise has zero average, $\Avg{\xi(t)} = 0$, and is essentially white: $\Avg{\xi(t)\xi(t')} \sim \delta(t-t')$.
Here the angle brackets indicate averaging over noise realizations.
The conservation of total nuclear spin is ensured by taking the noise to have the following spatial correlations on scales larger than the atomic lattice spacing:
\begin{equation}
\label{eq:Xi_stats}\Avg{\xi(\vec{x},t)\xi(\vec{x}',t')} = -\eta D\nabla^2\delta(\vec{x}-\vec{x}')\delta(t- t'), 
\end{equation}
where the proportionality constant $\eta$ will be fixed below to ensure the correct RMS value of the Overhauser difference field in equilibrium.
The units of $\eta$ are $[\hbar^2/\,{\rm Volume}]$.

Fourier transforming Eq.~(\ref{eq:Diff_realspace}), we obtain an independent differential equation for each nuclear spin mode $\tilde{I}_{\vec{q}}$, labeled by the 3D wave vector $\vec{q}$:
\begin{equation}
\label{eq:Diff_fourierspace} \partial_t \tilde{I}_{\vec{q}}(t) = -Dq^2 \tilde{I}_{\vec{q}} + \tilde{\xi}_{\vec{q}}(t).
\end{equation}
The Fourier modes $\tilde{\xi}_{\vec{q}}(t)$ of the noise field satisfy
\begin{equation}
\label{eq:Xi_stats_q} \Avg{\tilde\xi_{\vec{q}}(t)\tilde\xi_{\vec{q}'}(t')} =  \eta Dq^2 (2\pi)^3   \delta(\vec{q}+\vec{q}')\delta(t- t').
\end{equation}

The differential equation (\ref{eq:Diff_fourierspace}) has the formal solution 
\begin{equation}
\tilde{I}_{\vec{q}}(t) = \tilde{I}_{\vec{q}}(t_0) e^{-Dq^2 (t-t_0)} + \int_{t_0}^t dt'\, e^{-Dq^2(t-t')}\tilde{\xi}_{\vec{q}}(t').
\end{equation}
\begin{widetext}
Using the explicit form for $\tilde{I}_{\vec{q}}(t)$ above, along with Eq.~(\ref{eq:Xi_stats_q}), we obtain the correlation functions for the nuclear spin field:
\begin{eqnarray}
 \Avg{\tilde{I}_{\vec{q}}(t)\tilde{I}_{\vec{q}'}(t')}
 = e^{-Dq^2 (t + t'- 2t_0)}\left[\Avg{\tilde{I}_{\vec{q}}(t_0)\tilde{I}_{\vec{q}'}(t_0)} - \eta (2\pi)^3\delta(\vec{q} + \vec{q}')\right] + \eta (2\pi)^3\delta(\vec{q} + \vec{q}')e^{-Dq^2|t -t'|},
\end{eqnarray}
\end{widetext}
where in the second line we have used Eq.~(\ref{eq:Xi_stats_q}).
If the initial state $I(t_0)$ is drawn from the (stationary) equilibrium distribution, $\Avg{\tilde{I}_{\vec{q}}(t_0)\tilde{I}_{\vec{q}'}(t_0)} = \Avg{\tilde{I}_{\vec{q}}(t)\tilde{I}_{\vec{q}'}(t)}$, then we find
\begin{equation}
\label{eq:I_corr}\Avg{\tilde{I}_{\vec{q}}(t)\tilde{I}_{\vec{q}'}(t')} = \eta (2\pi)^3 \delta(\vec{q} + \vec{q}')e^{-Dq^2|t -t'|}.
\end{equation}
In position space, the equilibrium Overhauser field fluctuations are thus uncorrelated:
\begin{equation}
\label{eq:I_corr_x}\Avg{I(\vec{x},t)I(\vec{x}',t)} = \eta\, \delta(\vec{x}-\vec{x}').
\end{equation}

We now use the results above for the correlation function of the nuclear spin field to calculate the noise correlations of the Overhauser difference field, $\Delta B_z(t)$.
The correlation function $\Avg{\Delta B_z(t) \Delta B_z(t')}$ is straightforward to evaluate using Eqs.~(\ref{eq:dBzFourier}) and (\ref{eq:I_corr}):
\begin{equation}
\Avg{\Delta B_z(t) \Delta B_z(t')} 
\label{eq:dBzCorr}= \eta \frac{\hbar^2 A^2_0}{(g_*\mu_{\rm B})^2} \int \frac{d^3q}{(2\pi)^3}\, |\Delta\tilde\rho_{\vec{q}}|^2\,e^{-Dq^2|t -t'|},
\end{equation}
where we have used $\tilde{\rho}_{-\vec{q}} = \tilde{\rho}^*_{\vec{q}}$.

We set the value of $\eta$ by demanding that the equilibrium RMS Overhauser field fluctuations should match the measured value:
\begin{equation}
 \label{eq:dB2Eq} \sigma^2_{\Delta B} \equiv \Avg{\Delta B^2_z}_{\rm eq} = \eta \frac{\hbar^2 A^2_0}{(g_*\mu_{\rm B})^2} \int \frac{d^3q}{(2\pi)^3}\, |\Delta\tilde\rho_{\vec{q}}|^2.
\end{equation}
To evaluate $|\Delta \tilde{\rho}_{\vec{q}}|^2$ in the integrand above, we must specify a particular form for the electron density profiles in the two dots.
For simplicity we take the densities in the two dots to be Gaussian, centered at positions $\vec{x}_L = (x_L, 0, 0)$ and $\vec{x}_R = (x_R, 0, 0)$:
\begin{equation}
\label{eq:wavefunction}|\psi_\alpha(\vec{x})|^2 = \frac{e^{-[(x-x_\alpha)^2 + y^2]/(2\sigma_\perp^2)}}{2\pi\sigma_\perp^2}\frac{e^{-z^2/(2\sigma_z^2)}}{\sqrt{2\pi\sigma^2_z}},
\end{equation}
where $\alpha = L,R$. 
Setting $x_L = -d/2,\ x_R = d/2$ and taking the Fourier transform of the electron density in Eq.~(\ref{eq:wavefunction}), we obtain
\begin{equation}
|\Delta\tilde\rho_{\vec{q}}|^2 = 4\sin^2 (q_x d/2)\,e^{- (\sigma_z^2q_z^2\, +\, \sigma_\perp^2 q_x^2\, +\, \sigma_\perp^2 q_y^2)}.
\end{equation}

Rewriting $4\sin^2 (q_x d/2) = 2(1 - \cos q_xd)$ and substituting into Eq.~(\ref{eq:dB2Eq}) gives
\begin{eqnarray}
 \Avg{\Delta B^2_z}_{\rm eq}
&=&  \frac{\hbar^2 A^2_0}{(g_*\mu_{\rm B})^2}\frac{\eta}{4\pi^{3/2} \sigma_z\sigma_\perp^2}\left(1 - e^{-d^2/4\sigma_\perp^2}\right).
\end{eqnarray}
To cast the result into a more convenient form, we define the effective number of spins $N_\alpha$ in dot $\alpha = L,R$ via $N_\alpha^{-1} = v_0\int d^3x\, |\psi_\alpha(\vec{x})|^4$, where $v_0$ is the unit cell volume (see above).
For the wave functions in Eq.~(\ref{eq:wavefunction}) we have $N_\alpha = 8\pi^{3/2}\sigma_z\sigma^2_\perp/v_0 \equiv N$.
Letting $E_N = (g_*\mu_{\rm B})\sqrt{\Avg{\Delta B^2_z}_{\rm eq}}$, we have
\begin{equation}
\label{eq:Eq_Fluctuations}
E^2_N = \frac{2}{N}\cdot\frac{v_0 \eta}{\hbar^2}\cdot\left(\frac{\hbar^2 A_0}{v_0}\right)^2\cdot\left(1 - e^{-d^2/4\sigma_\perp^2}\right).
\end{equation}
In this form, the $\sqrt{N}$ dependence of the RMS Overhauser field fluctuations is explicitly displayed. 

The oscillations observed in Fig.~2 of the main text reveal the {\it magnitude} of the gradient, but do not yield information about its {\it sign}.
Therefore the correlation function in Eq.~(\ref{eq:dBzCorr}), which depends on both the magnitude and sign of $\Delta B(t)$, is not of direct relevance.
Instead, we compute the experimentally-relevant noise correlations and power spectrum for $\Delta B^2_z(t)$. 
Assuming that the noise is Gaussian, which is justified by the fact that the continuous nuclear spin field $I(\vec{x},t)$ is produced by a large density of randomly polarized individual spins, the fourth-order correlators that appear in the expression for $\Avg{\Delta B^2_z(t)\Delta B^2_z(t')}_c = \Avg{\Delta B^2_z(t)\Delta B^2_z(t')}_c - \Avg{\Delta B^2_z(t)}^2$ can be factorized.
This gives:
\begin{equation}
\label{eq:dBz2Corr2} \Avg{\Delta B^2_z(t) \Delta B^2_z(t')}_c 
 =  \frac{2\eta^2\hbar^4 A^4_0}{(g_*\mu_{\rm B})^4} 
\left[\int\!\! \frac{d^3q}{(2\pi)^3}|\Delta\tilde\rho_{\vec{q}}|^2 e^{-Dq^2|t - t'|}\right]^2\!\!\!.
\end{equation}

The integral in Eq.~(\ref{eq:dBz2Corr2}) is very similar to the one evaluated above to compute the RMS nuclear field.
Again using the electronic density profile, Eq.~(\ref{eq:wavefunction}), we get
\begin{widetext}
\begin{equation}
\label{eq:CorrelationInt}
C(t-t') \equiv \int \frac{d^3q}{(2\pi)^3}|\Delta\tilde\rho_{\vec{q}}|^2 e^{-Dq^2|t - t'|} \ =\ 
\frac{1 - e^{-\frac14 d^2/(\sigma_\perp^2  + D|t-t'|)}}{4\pi^{3/2}(D|t-t'| + \sigma_\perp^2)\sqrt{D|t-t'| + \sigma_z^2}}.
\end{equation}
Substituting back into Eq.~(\ref{eq:dBz2Corr2}), we get
\begin{equation}
\label{eq:AutoCorrdB2}
\Avg{\Delta B^2_z(t) \Delta B^2_z(t')}_c = \frac{\eta^2}{8\pi^3} \frac{\hbar^4 A^4_0}{(g_*\mu_{\rm B})^4}\frac{\left(1 - e^{-\frac14 d^2/(\sigma_\perp^2  + D|t-t'|)}\right)^2}{(D|t-t'| + \sigma_\perp^2)^2\,(D|t-t'| + \sigma_z^2)}. 
\end{equation}
\end{widetext}

The autocorrelation function in Eq.~(\ref{eq:AutoCorrdB2}) above was used for fitting the experimentally obtained power spectral densities for $(\Delta B_\parallel)^2$ in Fig.~2c.
The geometric parameters were taken from the lithographic dimensions of the device, and known growth parameters of the heterostructure: $d = 150$ nm, $\sigma_\perp = 40$ nm, and $\sigma_z = 7.5$ nm.
The diffusion constant $D$ and the equilibrium nuclear field fluctuations $E_N = g^*\mu_B\sigma_{\Delta B}$, see Eqs.~(\ref{eq:dB2Eq}) and (\ref{eq:Eq_Fluctuations}), were taken as fit parameters.
The extracted values were $D = 33$ nm$^2/s$ and $\sigma_{\Delta B} = 6.0$ mT. 

\subsection{Correlations for fixed separation time}

In Fig.~3 of the main text, power spectral densities for measurements of the singlet return probability with fixed separation time are shown.
This type of measurement was employed previously by Reilly and coworkers \cite{Reilly2008}.
For separation time $t_S$, the singlet return probability $P_S$ is given by
\begin{equation}
\label{eq:PSDef}P_S(t) = \frac12 {\rm Re}\left[1 + e^{i(g_*\mu_B/\hbar)\Delta B_z(t)t_S}\right].
\end{equation}
Here we assume that $B_z(t)$ is frozen on the timescale of one experiment, but its value may change from run to run.
Averaging over Gaussian fluctuations gives
\begin{equation}
\Avg{P_S(t)} = \frac12\left[1 + e^{-\frac12 (g_*\mu_B/\hbar)^2\Avg{\Delta B_z^2}t^2_S}\right].
\end{equation}

Using Eq.~(\ref{eq:PSDef}), the autocorrelation function $\Avg{P_S(t + \Delta t)P_S(t)} - \Avg{P_S(t)}^2$ then follows (as also found in Ref.~\cite{Reilly2008}):
\begin{widetext}
\begin{equation}
\Avg{P_S(t + \Delta t)P_S(t)} - \Avg{P_S(t)}^2 = \frac14 e^{-(g_*\mu_B/\hbar)^2\Avg{\Delta B^2_z}t_S^2}\left[\cosh \left( (g_*\mu_B/\hbar)^2\Avg{\Delta B_z(t + \Delta t) B_z(t)}t_S^2 \right) - 1 \right].
\end{equation}
\end{widetext}
The quantity $\Avg{\Delta B_z^2}$ in the exponent was calculated above, see Eq.~(\ref{eq:dB2Eq}) and below.
Furthermore, the correlation function $C(t - t') = \frac{(g_*\mu_B)^2}{\eta\hbar^2A_0^2} \Avg{\Delta B_z(t + \Delta t) B_z(t)}$ was calculated in Eq.~(\ref{eq:CorrelationInt}).
Thus we obtain the autocorrelation function for the singlet return probability in experiments with fixed separation time, used for fitting the data in Fig.~3 of the main text. 

\section{6. Decoherence of the qubit subjected to the transverse Overhauser noise}

\begin{figure}[tb]
	\includegraphics[scale=.9]{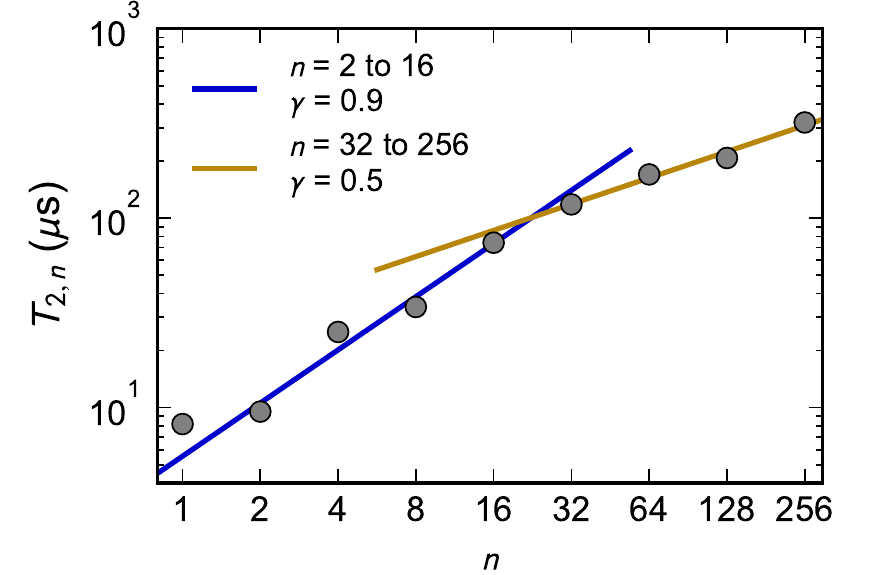}
	\caption{Scaling of the extracted coherence decay envelope $\Tn$ with $n$ at external magnetic field $B=0.75$~T. Solid blue and yellow lines indicate fits of the power law $\propto n^\gamma$ to data in the indicated range.}
	\label{figS4}
\end{figure}

For two electrons in the $(1,1)$ charge state of a double quantum dot, the effective pure dephasing spin Hamiltonian coupling the electron spins to the nuclei is given by \cite{Neder2011}
\begin{equation}
\hat{H}_{\mathrm{SN}} = \sum_{i=L,R}\left( \hat{h}^{i}_{z} + \frac{(\hat{h}^{i}_{x})^2 + (\hat{h}^{i}_{y})^2}{\Omega} \right )  \hat{S}^i_z \,\, \label{eq:HSN}
\end{equation}
where $\hat{h}^{i}_{a}$ with $a \! =\! x$, $y$, and $z$ are the operators of Overhauser field components, given by
$\hat{h}^{i}_{a} = \sum_{k} A^{i}_{k} \hat{I}^{a}_{k}$,
in which $\hat{I}^{a}_{k}$ are the spin operators of $k$-th nucleus and  $A^{i}_{k} = \mathcal{A}_{\alpha[k]} |\Psi_{i}(\mathbf{r}_{k})|^2$ (with $\Psi_{i}(\mathbf{r})$ being the envelope wavefunction of the electron in dot $i$, $\mathcal{A}_{\alpha[k]}$ the hyperfine interaction energy for nucleus of species $\alpha$, and $\mathbf{r}_{k}$ the position of $k$-th nucleus) are the hyperfine couplings of the $k$-th nucleus to the electron in dot $i$.

When the exchange interaction between the two electrons is strongly suppressed due to a large barrier height separating the two potential minima \cite{Martins2016}, the overlap between the $\Psi_{L}(\mathbf{r})$ and $\Psi_{R}(\mathbf{r})$ functions is negligible, and every contributing nucleus is coupled only to one electron, residing either in $L$ or $R$ dot. The Hamiltonian (\ref{eq:HSN}) is then a sum of two commuting terms, each pertaining to another dot. We also assume that the nuclear density matrices in the two dots are uncorrelated, and that the $L$ and $R$ dots have the same size and shape. 
The singlet return probability $P_{S}(T)$ is then given by
\begin{equation}
P_{S}(T) = \frac{1}{2} + \frac{1}{2}|W(T)|^2 \,\, ,  \label{eq:PSW}
\end{equation}
where $W(T)$ is the coherence function of a single spin in one of the QDs (i.e.~an off-diagonal element of its density matrix normalized to unity), calculated for the respective dynamical decoupling sequence. 

Although the longitudinal and transverse Overhauser field operators do not strictly commute, their commutator is $\! \sim \sigma/N$, where $\sigma$ is the rms of the Overhauser field and $N$ is the number of nuclei appreciably interacting with the electron \cite{Neder2011}. In the following we use a semiclassical approach to dynamics of large nuclear bath, and neglect this commutator. The decoherence function can then be written as
\begin{equation}
W(T) \approx W_{z}(T)W_{\perp}(T) \,\, ,
\end{equation}
in which $W_{z}(T)$ is the contribution to decoherence that originates from the first (longitudinal) term in Eq.~(\ref{eq:HSN}), while $W_{\perp}(T)$ is the contribution due to the second term (quadratic in transverse Overhauser operators). 

The Hamiltonian of the nuclei is the sum of a Zeeman term, a quadrupolar splitting term, and a dipole-dipole interaction term:
\begin{align}
\hat{H}_{\mathbf{N}} & =  \hat{H}_{Z} +  \hat{H}_{Q} + \hat{H}_{D} = \sum_{k}\omega_{k}\hat{I}^{z}_{k} + \sum_{j}q_{k}(I^{z}_{k})^2  \nonumber \\
&  + \sum_{k>l}b_{kl}(\hat{I}^{+}_{k}\hat{I}^{-}_{l} + \hat{I}^{-}_{k}\hat{I}^{+}_{l} - 2 \hat{I}^{z}_{k}\hat{I}^{z}_{l}) \,\, ,
\end{align}
where $\omega_{k}$  and $q_{k}$ are the Zeeman and quadrupolar splittings of the $k$-th nucleus respectively, and $b_{kl}$ is the dipolar coupling between $k$ and $l$ nuclei. 

It is important to note that $W_{\perp}(T)$ calculated for echo or other dynamical decoupling sequences has noticeable time-dependence due to the presence of $\hat{H}_{Z}$ and $\hat{H}_{Q}$ terms involving only single nuclei. The characteristic oscillations of $W_{\perp}(T)$ for spin echo \cite{Cywinski2009,Cywinski2009a,Bluhm2011,Neder2011,Botzem2016} and for CPMG \cite{Malinowski2017} arise from the presence of three distinct nuclear Larmor precession frequencies in GaAs.
In contrast, the echo decay envelope was explained \cite{Bluhm2011,Neder2011,Malinowski2017,Botzem2016} by the presence of a quadrupolar-induced spread of effective fields $\delta B$, with $\delta B$ sufficiently large to dominate over broadening due to $\hat{I}^{z}_{k}\hat{I}^{z}_{l}$ dipolar interactions. Reported values are $\delta B \! \approx \! 0.3$ mT in  \cite{Bluhm2011} and  $\delta B \! \approx \! 1$ mT in \cite{Malinowski2017,Botzem2016}. On the other hand, the decay of $W_z(T)$ is solely due to the dipolar $\hat{I}^{-}_{k}\hat{I}^{+}_{l}$  flip-flop term, which does not commute with $\hat{h}^{z}$ and thus leads to dynamics of the longitudinal Overhauser field. 

We calculate $W_{\perp}(T)$ using a semiclassical theory \cite{Neder2011}, in which the averages of products of any number $\hat{h}^{2}_{x,y}$ are evaluated to the lowest order in $1/N$ expansion \cite{Cywinski2009,Cywinski2009a}. Following \cite{Cywinski2009a,Neder2011} we define a $\mathcal{T}$-matrix, the components of which are given by
\begin{align}
\mathcal{T}_{kl}(T) & = \frac{2}{3}I(I+1)\sqrt{N_{k}N_{l}} \frac{A_{k}A_{l}}{2\Omega}\int_{0}^{T} f(t') e^{i\omega_{kl}t'} \nonumber\\ 
& \times \cos \left( \int_{0}^{t'}f(t'')A_{kl}\mathrm{d}t'' \right )  \,\, \label{eq:Tkl}
\end{align}
where $k(l)$ labels the group of $N_{k(l)}$ nuclei having (approximately) common values of hf coupling $A_{k(l)}$ and Zeeman splitting $\omega_{k(l)}$, $f(t')$ is the time-domain filter function corresponding to the given pulse sequence ($f(t')$ is nonzero for $t' \in [0,T]$, and it changes between $1$ and $-1$ value at times at which $\pi$ pulses are applied), $I$ is the length of individual nuclear spins (assumed to be the same for all spins, as is the case for GaAs, for which $I\! =\! 3/2$), and $\omega_{kl}\! = \! \omega_{k}-\omega_{l}$, $A_{kl} \! =\! (A_{k}-A_{l})/2$. We have then \cite{Cywinski2009,Cywinski2009a,Neder2011}
\begin{equation}
W_{\perp}(T) \! =\! \frac{1}{\mathrm{det}[1+i\mathcal{T}(T)]} = \exp \left [ \sum_{k=1}^{\infty} \frac{(-i)^k}{k}R_{k}(T) \right ] \,\, ,
\end{equation}
where $R_{k}(T) = \mathrm{Tr}[\mathcal{T}^{k}(T)]$.

For a spin echo sequence of length $T\! = \! \tau$ the $\mathcal{T}$-matrix is given by \cite{Cywinski2009,Cywinski2009a,Neder2011}
\begin{align}
\mathcal{T}^{\mathrm{SE}}_{kl}(\tau) & = \frac{\bar{b}_{k}\bar{b}_{l}}{\Omega}  \frac{-i\omega_{kl}}{\omega^{2}_{kl}-A^{2}_{kl}}  
\left( \cos \frac{\omega_{kl}\tau}{2} - \cos \frac{A_{kl}\tau}{2} \right) e^{i\omega_{kl}\tau/2} \,\, , \label{eq:TklSE}
\end{align}
while for a CMPG sequence with even number of pulses $n$ and interpulse spacing $\tau$ we have \cite{Malinowski2017}
\begin{align}
\mathcal{T}^{\mathrm{CP,n}}_{kl}(T=n\tau) & = \frac{\bar{b}_{k}\bar{b}_{l}}{\Omega}  \frac{\omega_{kl}}{\omega^2_{kl}-A^{2}_{kl}} \nonumber 
\frac{\cos \frac{\omega_{kl}\tau}{2} - \cos \frac{A_{kl}\tau}{2}}{\cos \frac{\omega_{kl}\tau}{2}}  \nonumber \\
& \times \sin \frac{\omega_{kl}n\tau}{2} e^{i\omega_{kl}n\tau/2} \,\, . \label{eq:TklCP}
\end{align}
where $\bar{b}_{k} \! =\! \sqrt{\frac{2}{3}I(I+1)N_{k}}A_{k}$ is the rms strength of the Overhauser field arising from $N_{k}$ spins, all spins having Knight shift $A_{k}$. 

It is crucial now to recognize the distinct roles of two kinds of contributions to the $\mathcal{T}$-matrix: the heteronuclear ones, in which $\omega_{k}$ and $\omega_{l}$ correspond to distinct nuclear isotopes (i.e.~$^{69}$Ga, $^{71}$Ga, and $^{75}$As in the case of GaAs, labeled by $\alpha \! =\! 1,2,3$), so that $\omega_{kl} \! \approx \! \omega_{\alpha\beta}$, and the homonuclear ones, in which $\omega_{k}$ and $\omega_{l}$ correspond to groups of nuclei of the same isotope, so that $\omega_{kl} \! \ll \! \omega_{\alpha\beta}$. 
The former terms govern the presence of characteristic oscillations of $W_{\perp}(t)$ in echo \cite{Bluhm2011,Botzem2016,Malinowski2017} and CPMG case \cite{Malinowski2017}, while the latter smooth these oscillations in CPMG case and, more importantly,  lead to an irreversible decay of the signal. Note that the homonuclear terms are nonzero in the presence of intra-species spread of nuclear splittings, thereby contributing to low- and intermediate-frequency noise. 

Note that the dependence of $\mathcal{T}_{kl}$ on Knight shift differences $A_{kl}$ is negligible in experimentally relevant range of parameters for both homonuclear and heteronuclear terms. For QD with number of nuclei $N \! \approx \! 10^{6}$ we have $A_{kl}\tau \! \ll \! 1$ for $\tau \! \ll \! 20$ $\mu$s. For heteronuclear terms we can then put $A_{kl}\! =\! 0$ in Eqs.~(\ref{eq:TklSE}) and (\ref{eq:TklCP}) provided that $\omega_{\alpha\beta} \! \gg \! A_{kl}$ (which is fulfilled for $B_{\text{ext}} \! > \! 100$ mT in dots considered here), while for homonuclear terms for isotope $\alpha$ we have to assume that $\omega_{kl}\tau \! \ll \! 1$ (which is easily fulfilled for considered values of $\tau$ with $\delta B \! \approx \! 1$ mT used here). We can then perform the calculation by dividing the nuclei of each isotope into $K$ groups, each having the same $A_{k}$ (equal to the typical hf coupling for a give isotope), value of $\omega_{k}$ taken from $[\omega_{\alpha}-5\sigma_{\alpha},\omega_{\alpha}+5\sigma_{\alpha}]$ range, and $N_{k}$ taken from a Gaussian distribution:
\begin{equation}
N_{k\in \alpha} = N_{\alpha} \frac{1}{\sqrt{2\pi}\sigma_{\alpha}} \exp\left (-\frac{(\omega_{k}-\omega_{\alpha})^2}{2\sigma^{2}_{\alpha}} \right) \,\, ,
\end{equation}
in which $\sigma_{\alpha}$ is the rms value of the nuclear splitting due to a spread $\delta B$ of the effective field experienced by the nuclei. The calculations converge on timescales relevant for measurements presented in this paper for $K \! < \! 100$ for $n\! =\! 256$ pulses (and for smaller $K$ for lower numbers of pulses).

\begin{figure}
	\centering
	\includegraphics[scale=.9]{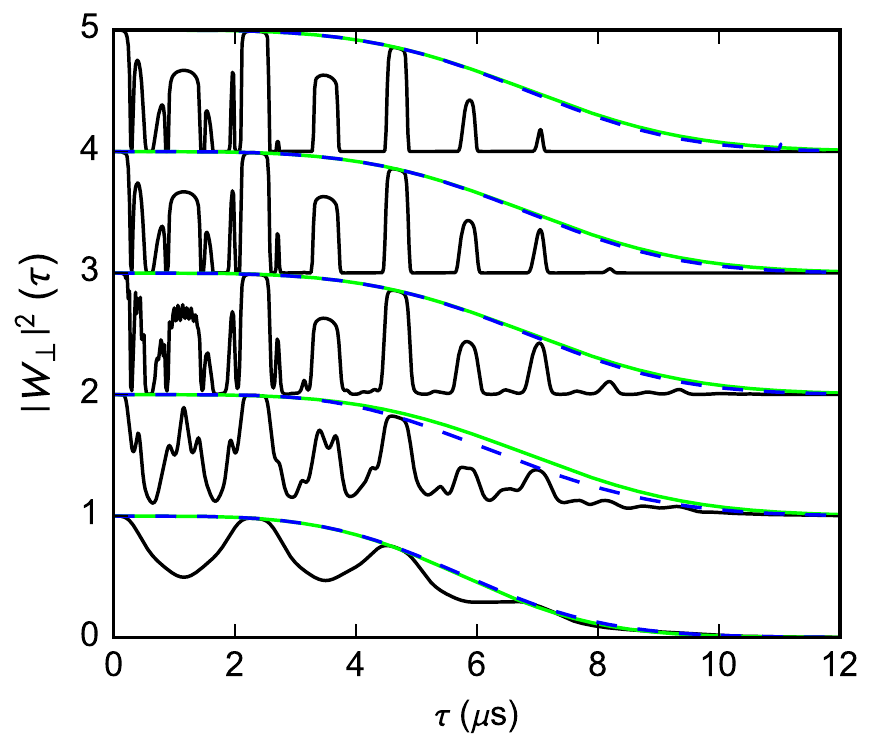}
	\caption{Two-spin decoherence function $|W_{\perp}(T=n\tau)|^2$ calculated using the $T$-matrix approach for $n\! = \! 1$, $4$, $16$, $64$, and $256$ (from bottom to top). Black solid lines are the exact calculation, green lines are the homonuclear-only result, and dashed lines correspond to analytical approximation to the homonuclear result from Eq.~(\ref{eq:Whoman}).}
	\label{figS5}
\end{figure}

Examples of results for $n\! = 1$, $4$, $16$, $64$, and $256$ are shown in Fig.~\ref{figS5}. In the calculations we have used the values of $A_{\alpha} \! = \! 2\mathcal{A}_{\alpha}/N$, with $\mathcal{A}_{\alpha} \! =\! 35.9$, $45.9$, and $42.9$ $\mu$eV, and $\omega_{\alpha} \! = \! -42.1$, $-53.6$, and $-30.1$ neV at $1$ T field corresponding to $^{69}$Ga, $^{71}$Ga, and $^{75}$As, respectively (note that $N$ here is the number of nuclei, while in \cite{Cywinski2009a} it denoted the number of unit cells, i.e.~twice the number of nuclei). The experimental results presented in the main text are best fit by $N \! =\! 9 \times 10^{5}$ in each dot, and effective broadening $\sigma_{\alpha}$ corresponding to $1$ mT field, and these are the values used in Fig.~\ref{figS5}.

The key observation is that the calculated $W_{\perp}(T)$ can be very well approximated by a product of decoherence functions calculated while keeping {\it only} the homonuclear terms and {\it only} the heteronuclear ones:
\begin{equation}
W_{\perp}(T) \approx W_{\perp,\text{het}}(T) \times W_{\perp,\text{hom}}(T)   \,\, .
\end{equation}
The heteronuclear term is responsible for large-amplitude oscillations of the signal, while the homonuclear term gives a decay envelope of the coherence signal, see Fig.~\ref{figS5}.
Furthermore, the homonuclear contribution can be approximated very well (at least on timescale of the signal decay) by a simple solution obtained using a bimodal approximation to the distribution of $\omega_{k}$ frequencies of nuclei of each species (first used for spin echo case in \cite{Neder2011}). In this approximation we have
\begin{equation}
W_{\perp,\text{hom}}(T) \approx \prod_{\alpha}\frac{1}{1+(T/t^{(n)}_{\alpha})^4} \,\, ,\label{eq:Whoman}
\end{equation}
where for echo we have 
\begin{equation}
t^{(1)}_{\alpha} = \frac{2\sqrt{2}}{\bar{b}_{\alpha}} \sqrt{\frac{\Omega}{\sigma_{\alpha}}} \,\, ,\label{eq:tSE}
\end{equation}
while for CPMG sequence with even $n$ we obtain
\begin{equation}
t^{(n)}_{\alpha} = 2^{1/4} n t^{(1)}_{\alpha} \,\, .
\end{equation}
This is the main result here: when decoherence due to tranvserse Overhauser field fluctuations is dominated by the homonuclear contribution, the characteristic coherence half-decay time $T_{2}$ (defined by $W_{\perp,\text{hom}}(T_{2})\! =\! 1/e$) scales {\textit {linearly}} with the number of pulses $n$, i.e.~we have $T_{2}\propto n^{\gamma_{\perp}}$ with $\gamma_{\perp}\!=\! 1$.

\end{document}